\documentclass{article}

\usepackage[T1]{fontenc}
\usepackage[utf8]{inputenc}
\usepackage[english]{babel}
\usepackage[backend = biber, style = alphabetic, sorting = ynt, citestyle = authoryear]{biblatex} 
\DeclareCiteCommand{\cite}
{\usebibmacro{prenote}}
{\usebibmacro{citeindex}%
	\printtext[bibhyperref]{\usebibmacro{cite}}}
{\multicitedelim}
{\usebibmacro{postnote}}

\DeclareCiteCommand*{\cite}
{\usebibmacro{prenote}}
{\usebibmacro{citeindex}%
	\printtext[bibhyperref]{\usebibmacro{citeyear}}}
{\multicitedelim}
{\usebibmacro{postnote}}

\DeclareCiteCommand{\parencite}[\mkbibparens]
{\usebibmacro{prenote}}
{\usebibmacro{citeindex}%
	\printtext[bibhyperref]{\usebibmacro{cite}}}
{\multicitedelim}
{\usebibmacro{postnote}}

\DeclareCiteCommand*{\parencite}[\mkbibparens]
{\usebibmacro{prenote}}
{\usebibmacro{citeindex}%
	\printtext[bibhyperref]{\usebibmacro{citeyear}}}
{\multicitedelim}
{\usebibmacro{postnote}}

\DeclareCiteCommand{\footcite}[\mkbibfootnote]
{\usebibmacro{prenote}}
{\usebibmacro{citeindex}%
	\printtext[bibhyperref]{ \usebibmacro{cite}}}
{\multicitedelim}
{\usebibmacro{postnote}}

\DeclareCiteCommand{\footcitetext}[\mkbibfootnotetext]
{\usebibmacro{prenote}}
{\usebibmacro{citeindex}%
	\printtext[bibhyperref]{\usebibmacro{cite}}}
{\multicitedelim}
{\usebibmacro{postnote}}

\DeclareCiteCommand{\textcite}
{\boolfalse{cbx:parens}}
{\usebibmacro{citeindex}%
	\printtext[bibhyperref]{\usebibmacro{textcite}}}
{\ifbool{cbx:parens}
	{\bibcloseparen\global\boolfalse{cbx:parens}}
	{}%
	\multicitedelim}
{\usebibmacro{textcite:postnote}}

\usepackage[ruled]{algorithm2e}

\usepackage{amsmath}
\usepackage{latexsym}
\usepackage{amsthm}
\usepackage{amsfonts}
\usepackage{amssymb}
\usepackage{mathrsfs}
\usepackage{dsfont}
\usepackage{bbm}
\usepackage{graphicx}
\graphicspath{{plots/}}

\usepackage{longtable}
\usepackage{parskip}
\usepackage{enumerate}
\usepackage{hyperref}
\usepackage{wrapfig}
\usepackage{multirow}
\usepackage{hhline}
\usepackage{tabularx}
\usepackage{adjustbox}
\usepackage{pdflscape}
\usepackage{rotating}
\usepackage{array}
\usepackage{tikz}
\usepackage{float}
\usepackage{fancyvrb}
\usepackage{csquotes}
\usepackage{verbatim}
\usepackage{color}
\usepackage{soul}
\usepackage{booktabs}
\usepackage[hmargin=1in,vmargin=1in]{geometry}
\definecolor{blue}{HTML}{0000EE}
\definecolor{black}{HTML}{000000}
\hypersetup{
	linkcolor  = blue,
	citecolor  = blue,
	urlcolor   = black,
	colorlinks = true,
}
\usepackage[
open,
openlevel=2,
atend,
numbered
]{bookmark}

\usepackage{cleveref}

\crefformat{section}{\S#2#1#3} 
\crefformat{subsection}{\S#2#1#3}
\crefformat{subsubsection}{\S#2#1#3}
\usepackage{afterpage}

\def\RR{\mathds{R}}

\def\EE{\mathds{E}}

\def\DDD{\mathscr{D}}

\def\NNN{\mathscr{N}}

\def\ind{\mathds{1}}

\newcommand{\cmmnt}[1]{\ignorespaces}

\numberwithin{equation}{section}

\newtheorem{thm}{Theorem}

\newtheorem{cond}{Condition}

\DeclareMathOperator*{\argmax}{arg\,max}

\allowdisplaybreaks

\begin{document}
	
	\title{Generalised Boosted Forests}
	\author{\textbf{Indrayudh Ghosal} \\ {\small \href{mailto:IG248@CORNELL.EDU}{\nolinkurl{IG248@CORNELL.EDU}}} \and \textbf{Giles Hooker} \\ {\small \href{mailto:GJH27@CORNELL.EDU}{\nolinkurl{GJH27@CORNELL.EDU}}} \and Department of Statistics and Data Science, Cornell University \\ Ithaca, NY 14850, USA}
	\maketitle
	
	\begin{abstract}
		This paper extends recent work on boosting random forests to model non-Gaussian responses. Given an exponential family $\EE[Y|X] = g^{-1}(f(X))$ our goal is to obtain an estimate for $f$. We start with an MLE-type estimate in the link space and then define generalised residuals from it. We use these residuals and some corresponding weights to fit a base random forest and then repeat the same to obtain a boost random forest. We call the sum of these three estimators a \textit{generalised boosted forest}. We show with simulated and real data that both the random forest steps reduces test-set log-likelihood, which we treat as our primary metric. We also provide a variance estimator, which we can obtain with the same computational cost as the original estimate itself. Empirical experiments on real-world data and simulations demonstrate that the methods can effectively reduce bias, and that confidence interval coverage is conservative in the bulk of the covariate distribution. 
	\end{abstract}

	\section{Introduction} 
	
	This paper extends recent work on boosting random forests [\cite{ghosal2020boosting}] to model non-Gaussian responses.  Ensembles of trees have become one of the most successful general-purpose machine learning methods, with the advantage of being both computationally efficient and having few tuning parameters that require human intervention [\cite{breiman2001random}, \cite{friedman2001elements}].  Ensembles are made of multiple weak learners -- most commonly based on tree-structured models -- that are combined by one of two broad strategies: bagging or boosting.  Bagging methods, in particular random forests [\cite{breiman2001random}], combine trees that are learned in parallel using the same processes: usually differing in the data samples given to the trees and any randomisation processes used in obtain them. In contrast, boosting methods, such gradient boosted forests [\cite{friedman2001greedy}], obtain trees in sequence: each tree used to improve the prediction of the current ensemble.
	
	\cite{ghosal2020boosting} proposed combining bagging and boosting by boosting a random forest estimator: a second random forest was built to predict the out of bag residuals of the first random forest in a regression setting. This approach can be traced back to \cite{breiman2001random} and achieves a near-universal improvement in test-set error on real-world problems. A significant motivation in \cite{ghosal2020boosting} was to extend recent results on uncertainty quantification for random forests to boosting-type methods. \cite{mentch2016quantifying} and \cite{wager2017estimation} provided confidence intervals for the predictions of random forests by employing a particular bagging structure, when each tree is given by a strict subsample of the dataset. This result is obtained by representing the ensemble as an infinite order incomplete $U$-statistic for which a central limit theorem can be obtained with the limiting variance calculated either by a direct representation [\cite{mentch2016quantifying}] or the Infinitesimal Jackknife [\cite{wager2017estimation}], see also \cite{zhou2019asymptotic}. \cite{ghosal2020boosting} extended this framework to represent the combination of sequentially-built random forests as a sum of two jointly-normal $U$-statistics for which an extension of the Infinitesimal Jackknife could be applied. 
	
	This paper builds on the ideas in \cite{ghosal2020boosting} to the generalised regression setting where the responses follow an exponential family distribution. We use binary responses for classification and count data in which the response follows a Poisson distribution as motivating examples.  In the general exponential family we assume that the expected value of the response is related to a function of the predictor but only through a suitable link function, i.e, $\EE[Y|X] = g^{-1}(f(X))$. When  restricted to a linear model $f(x) = x^\top \beta$, this follows the generalized linear model framework, but is replaced in this paper by the sum of sequentially-constructed random forests. We estimate this model via Newton-boosting for each random forest step, which is essentially a Newton-Raphson update in gradient boosting [\cite{friedman2001greedy}]. This type of updating has been used in other iterative methods, for example word-embedding as in \cite{baer2018exponential}. We first obtain an optimal constant, and then fit two random forests in sequence, each time using a weighted fit to Newton residuals. This results in the same random forest process being used for each step; see \cref{sec:gbfdefine} for details. 
	
	We also extend the variance estimates derived in \cite{ghosal2020boosting} to this generalized regression model. This includes an extension of the Infinitesimal Jackknife estimate of covariance to include accounting for the initial constant; see \cref{sec:varest}. In \cref{sec:gbftheory} we discuss some theoretical properties of our algorithm, namely its asymptotic normality and consistency of the variance estimate under suitable regularity conditions. Finally in \cref{sec:gbfresults} we empirically demonstrate the utility of our algorithm - \cref{sec:simresults} focuses on results from a simulated dataset and \cref{sec:realresults} demonstrates the performance of our algorithm on some datasets from the UCI database (\cite{lichman2013uci}). In both cases we measure performance mainly in terms of test-set log likelihood, although other related metrics are also considered.

	\section{Defining Generalised Boosted Forests}\label{sec:gbfdefine}
	
	We consider the generalised model $\EE[Y|X] = g^{-1}(f(X))$ where $g$ is an appropriate link function; in this paper we consider the specific cases of the logit function if $Y$ is binary, or the log function if $Y$ is count data. Given data $(Z_i)_{i=1}^n = (Y_i, X_i)_{i=1}^n$ our goal is to construct an estimate $\hat{f}$ for $f$. Note that in the generalized linear model literature [\cite{mccullagh1989generalized}] the estimate $\hat{f}$ is from the family of linear functions but all of our calculations below can apply to any family of functions. Throughout this paper we will refer to the estimate $\hat{f}$ and its properties to be in the "link" space, and we will refer to $g^{-1}(\hat{f})$ to be in the "response" space.
	
	We define the log-likelihood for the dataset $(Y_i, X_i)_{i=1}^n$ to be $\sum_{i=1}^n \ell(\eta_i; Y_i, X_i)$, where $\eta_i = f(X_i)$ is in the link space. For notational simplicity we will refer to the likelihood function for one point to be simply $\ell_i(\eta_i)$ without explicit reference to the dependence on the data $(Y_i, X_i)$. To estimate $\hat{f}$, first we will obtain an MLE-type initial value $\hat{\eta}_{MLE}^{(0)}$ given by
	$$
	\hat{\eta}_{MLE}^{(0)} = \argmax_{t} \sum_{i=1}^n \ell_i(t)
	$$
	
	To improve upon this estimate we will define "residuals" and then fit a function $\hat{f}^{(1)}$ with these residuals as responses. In \cite{ghosal2020boosting} we used the Gaussian regression setting with the model being $Y = f(X) + \epsilon$, where $\epsilon \sim N(0, \sigma^2)$. Then the residuals are given by $r_i = Y_i - \hat{f}(X_i)$, where $\hat{f}$ is some estimator of $f$. To generalise the definition of residuals to the non-Gaussian setting we will first note the relationship between residuals and the log-likelihood in the Gaussian setting. The log-likelihood in this case is given by $\sum_{i=1}^n \ell_i(\hat{f}(X_i)) = C - \sum_{i=1}^n \frac{(Y_i - \hat{f}(X_i))^2}{2\sigma^2}$, where $C$ is a constant. Then we see that $r_i \propto \ell_i'(\hat{f}(X_i))$. 
	
	Our goal now is to find a function $\hat{f}^{(1)}$ such that the log-likelihood has improved from its previous value while also utilising some notion of "residuals" as related to the derivative of the log-likelihood. So we must try to maximise $\sum_{i=1}^n \ell_i(\eta_i)$, i.e., minimise $\sum_{i=1}^n -\ell_i(\eta_i)$. Using a second order Taylor expansion on $-\ell_i$ we note that:	
	\begin{align*}
		\sum_{i=1}^n -\ell_i(\eta_i) &\approx \sum_{i=1}^n \left[ - \ell_i(\hat{\eta}_i^{(0)}) - \ell_i'(\hat{\eta}_i^{(0)}) \cdot (\eta_i - \hat{\eta}_i^{(0)}) - \ell_i''(\hat{\eta}_i^{(0)}) \cdot \frac{(\eta_i - \hat{\eta}_i^{(0)})^2}{2} \right] \\
		&= \sum_{i=1}^n -\ell_i''(\hat{\eta}_i^{(0)}) \left( \eta_i - \hat{\eta}_i^{(0)} + \frac{\ell_i'(\hat{\eta}_i^{(0)})}{\ell_i''(\hat{\eta}_i^{(0)})} \right)^2 + C,
	\end{align*}
	where $C$ is some constant. We can then approximate $\xi_i = \eta_i - \hat{\eta}_i^{(0)}$ using a random forest trained with predictors $X_i$, training responses $\ell_i'(\hat{\eta}_i^{(0)})/\big(-\ell_i''(\hat{\eta}_i^{(0)})\big)$ (the weighted residuals) and training weights $-\ell_i''(\hat{\eta}_i^{(0)})$. This sort of "weighted fit" is akin a Newton-Raphson update. Denote this base random forest by $\hat{f}^{(1)}$. Here we note that:
	\begin{itemize}
		\item This calculation holds for any initial estimate $\hat{f}^{(0)}(X_i) = \hat{\eta}_i^{(0)}$ but for the purposes of this paper the initial estimate $\hat{\eta}_i^{(0)} = \hat{\eta}_{MLE}^{(0)}$ is constant over $i$
		\item There may be different ways of implementing the training weights while fitting a random forest. This paper employs the \textsf{ranger} package in \textsf{R} where weights are used as the sampling weights of the datapoints given to each tree. Alternative approaches might include utilising the weights in a weighted squared error loss function to be used for the splitting criteria in construction of each tree. We will assume that these two procedures (and any procedure using reasonable re-weighting) are equivalent.
		\item In the Gaussian regression case the second derivatives $\ell_i''(\hat{\eta}_i^{(0)})$ are constant. In the non-Gaussian case we could also make the same assumption and simply fit random forests with the residuals $\ell_i'(\hat{\eta}_i^{(0)})$ as the response (with weights assumed to be 1 without loss of generality). Initial experiments (not shown) suggest that this procedure does not perform better than Newton weighting.
	\end{itemize}
	
	Once we have $\hat{\xi}_i^{(1)} = \hat{f}^{(1)}(X_i)$ we can also fit another random forest as a boosted step. Following similar calculations as above we will use predictors $X_i$, weighted residuals $\ell_i'(\hat{\eta}_i^{(0)}+\hat{\xi}_i^{(1)})/\big(-\ell_i''(\hat{\eta}_i^{(0)}+\hat{\xi}_i^{(1)})\big)$ as the training response and training weights $-\ell_i''(\hat{\eta}_i^{(0)}+\hat{\xi}_i^{(1)})$. Denote this boosted-stage random forest by $\hat{f}^{(2)}$.
	
	The final predictor for a test point $x$ will be given by 
	$$
	\hat{f}(x) = \hat{\eta}_{MLE}^{(0)} + \hat{f}^{(1)}(x) + \hat{f}^{(2)}(x)
	$$
	in the link space and by $g^{-1}(\hat{f}(x))$ in the response space. Similar to the Gaussian case discussed in \cite{ghosal2020boosting} we could also continue this boosting for more than one step; however we expect diminishing returns (and increasing computational burden) with further boosting steps as the size of the remaining signal decreases relative to the intrinsic variance in the data.  
	
	\section{Uncertainty Quantification}
	
	In this section we derive a variance estimate that can be calculated at no additional cost.	We also show that the variance estimate is consistent under some regularity conditions and motivate a central limit theorem. Finally we end the section with an algorithmic description of the generalised boosted forest.
	
	\subsection{Variance Estimation}\label{sec:varest}
	
	\paragraph{The Infinitesimal Jackknife}
	
	We use the Infinitesimal Jackknife (IJ) method to calculate the variance estimates of each stage and also for the final predictor (in the link space). In general \cite{efron1982jackknife} defines the Infinitesimal Jackknife for an estimate of the form $\hat{\theta}(P^0)$ where $P^0$ is the uniform probability distribution over the empirical dataset. First we consider the slightly more general estimate $\hat{\theta}(P^*)$ where $P^*$ is some probability distribution over the empirical dataset, i.e., it is a vector of the same length as the size of the dataset and has positive elements that add to 1. This is then approximated by the hyperplane tangent to the surface $\hat{\theta}(P^*)$ at the point $P^* = P^0$, i.e., $\hat{\theta}(P^*) \approx \hat{\theta}_{\text{TAN}}(P^*) = \hat{\theta}(P^0) + (P^*-P^0)^\top U$, where $U$ is a vector of the directional derivatives given by
	$$
	U_i = \lim_{\epsilon \to 0} \frac{\hat{\theta}(P^0 + \epsilon(\delta_i - P^0)) - \hat{\theta}(P^0)}{\epsilon}, i = 1,\ldots,n
	$$
	Under a suitable asymptotic normal distribution for $P^*-P^0$ we can obtain the variance for $\hat{\theta}_{\text{TAN}}(P^*)$ to be $\frac1{n^2} \sum_{i=1}^n U_i^2$. This is the IJ variance estimate for the estimator $\hat{\theta}(P^0)$.
	
	\paragraph{In the link space}
	
	Borrowing the same notation as above suppose $\hat{\eta}_{MLE}^{(0)} = \hat{\theta}(P^0)$, where $P^0$ is the uniform probability vector and its directional derivatives are given by $U_i$. Then the IJ variance estimate of $\hat{\eta}_{MLE}^{(0)}$ will be given by $\widehat{var}_{IJ}(\hat{\eta}_{MLE}^{(0)}) = \frac1{n^2} \sum_{i=1}^n U_i^2$. \cite{wager2014confidence} demonstrate that the directional derivatives for random forests are given by
	$$
	U'_i = n \cdot cov_b (N_{i,b}, T_b(x))
	$$
	where $N_{i,b}$ is the number of times the $i$\textsuperscript{th} datapoint is used in training the $b$\textsuperscript{th} tree, $T_b(x)$ is the prediction for a testpoint $x$ from the $b$\textsuperscript{th} tree, and the covariance is calculated over all trees in the forest. Thus for our predictor $\hat{f}$ the IJ variance estimate is given by
	\begin{align}
		\hat{V}(x) &= \frac1{n^2} \sum_{i=1}^n \left( U_i + n \cdot cov_b (N_{i,b}^{(1)}, T_b^{(1)}(x)) + n \cdot cov_b (N_{i,b}^{(2)}, T_b^{(2)}(x)) \right)^2 \nonumber \\
		&\qquad\qquad + \frac1B \big(var_b(T_b^{(1)}(x)) + var_b(T_b^{(2)}(x)) \big) \label{eq:varest}
	\end{align}
	where the second term accounts for the extra variability from the random forest being an incomplete U-statistic rather than a complete U-statistic [\cite{ghosal2020boosting}]. In \cref{sec:varcons} we show that this variance estimate is consistent under some regularity assumptions detailed in \cref{sec:gbftheory}. We would especially like to note one of the cross terms in the first term of the above expression given by $\frac1n \sum_{i=1}^n U_i \cdot cov_b (N_{i,b}^{(1)}, T_b^{(1)}(x))$ which is the IJ estimate of the covariance between the constant ($\hat{\eta}_{MLE}^{(0)}$) and the base stage random forest ($\hat{f}^{(1)}$). Similar cross-product terms will be estimates of co-variance between $\hat{\eta}_{MLE}^{(0)}$ (and $\hat{f}^{(1)}$) with the boosted stage random forest $\hat{f}^{(2)}$.
	
	\paragraph{Bias correction}
	
	The IJ variance estimate is known to have an upward bias, and so we will employ a correction term before further exploration. Our bias correction is inspired by \cite{zhou2019asymptotic}, namely the final formula for the variance estimate of a U-statistic in appendix E of that paper. Although note that we don't need to use the multiplicative factor because we already account for the variance between the trees in the second term of our variance estimate above \eqref{eq:varest}. Our final variance estimate is given by
	\begin{align}
		\hat{V}(x) &= \frac1{n^2} \sum_{i=1}^n \left( U_i + n \cdot cov_b (N_{i,b}^{(1)}, T_b^{(1)}(x)) + n \cdot cov_b (N_{i,b}^{(2)}, T_b^{(2)}(x)) \right)^2 \nonumber \\
		&\qquad\qquad + \frac1B \cdot \left(1 - \frac{n}{k}\right) \cdot \big(var_b(T_b^{(1)}(x)) + var_b(T_b^{(2)}(x)) \big) \label{eq:varestcorr}
	\end{align}
	where $k$ is the size of the subsamples (without replacement) for each tree in the random forests $\hat{f}^{(1)}$ and $\hat{f}^{(2)}$.
	
	\paragraph{Examples}
	
	The closed form of $U_i$ will depend on the type of response being used. For example we can calculate $U_i$ for two cases below:
	\begin{itemize}
		\item For the binomial response family ($g^{-1}$ is the logit function) our model is $Y_i \sim binomial(n_i, p_i)$ where $p_i = g^{-1}(f(X_i))$ and $n_i$ are given positive integers. So we estimate $\hat{\eta}_i^{(0)}$ as follows ($C$ is a constant)
		\begin{align*}
			\ell_i(t) &= y_i \log\left(\frac{e^t}{1+e^t}\right) + (n_i-y_i)\log\left(\frac1{1+e^t}\right) + C \\
			&= y_i(t - \log(1+e^t)) - (n_i-y_i)\log(1+e^t) + C \\
			&= y_i t - n_i\log(1+e^t) + C\\
			\implies \ell_i'(t) &= y_i - n_i \cdot \frac{e^t}{1+e^t}
		\end{align*}
		Then $\displaystyle\sum_{i=1}^n \ell_i'(t) = 0 \implies \displaystyle\sum_{i=1}^n y_i - \frac{e^t}{1+e^t} \cdot \displaystyle\sum_{i=1}^n n_i = 0 \implies \hat{\eta}_i^{(0)} = t = \log\left(\frac{\sum_{i=1}^n y_i}{\sum_{i=1}^n (n_i-y_i)}\right)$. Thus $\hat{\eta}_i^{(0)} = \hat{\theta}(P^0)$ where $P^0$ is the uniform probability vector and $\hat{\theta}(P) = \log\left(\frac{Y^\top P}{(N-Y)^\top P}\right)$ for a general probability vector $P$. So
		\begin{align*}
			U_i &= \lim_{\epsilon \to 0} \frac{\hat{\theta}(P^0 + \epsilon(\delta_i - P^0)) - \hat{\theta}(P^0)}{\epsilon} \\
			&= \lim_{\epsilon \to 0} \frac{\log(Y^\top (P^0 + \epsilon(\delta_i - P^0))) + \log((N-Y)^\top (P^0 + \epsilon(\delta_i - P^0)))}{\epsilon} \\
			&\qquad\qquad - \frac{\log(Y^\top P^0) + \log((N-Y)^\top P^0)}{\epsilon} \\
			&= \lim_{\epsilon \to 0} \frac{\log(Y^\top P^0 + \epsilon\cdot Y^\top(\delta_i - P^0))) - \log(Y^\top P^0)}{\epsilon} \\
			&\qquad\qquad - \frac{\log((N-Y)^\top P^0 + \epsilon\cdot (N-Y)^\top(\delta_i - P^0))) - \log((N-Y)^\top P^0)}{\epsilon} \\
			&= \frac{Y^\top (\delta_i - P^0)}{Y^\top P^0} - \frac{(N-Y)^\top (\delta_i - P^0)}{(N-Y)^\top P^0} \\
			&= \frac{y_i - \bar{y}}{\bar{y}} - \frac{(n_i - y_i) - (\bar{n} - \bar{y})}{\bar{n} - \bar{y}} \\
			&= \frac{\bar{n}y_i - n_i \bar{y}}{\bar{y}(\bar{n} - \bar{y})}
		\end{align*}
		
		\item In case where $Y$ is count data and we use a Poisson response family with our model being $Y_i \sim Poisson(\lambda_i)$, where $\lambda_i = \exp(f(X_i))$ and we get the following calculations for estimating $\hat{\eta}_i^{(0)}$ ($C$ is a constant)
		\begin{align*}
			\ell_i(t) &= -e^t + y_i \log(e^t) + C = y_i t - e^t + C \implies \ell_i'(t) = y_i - e^t \\
			\text{Then } \sum_{i=1}^n \ell_i'(t) &= 0 \implies \sum_{i=1}^n (y_i - e^t) = 0 \implies \hat{\eta}_i^{(0)} = t = \log(\bar{y})
		\end{align*}
		So we have $\hat{\eta}_i^{(0)} = \hat{\theta}(P^0)$ with $\hat{\theta}(P) = \log(Y^\top P)$ and thus
		\begin{align*}
			U_i &= \lim_{\epsilon \to 0} \frac{\hat{\theta}(P^0 + \epsilon(\delta_i - P^0)) - \hat{\theta}(P^0)}{\epsilon} \\
			&= \lim_{\epsilon \to 0} \frac{\log(Y^\top (P^0 + \epsilon(\delta_i - P^0))) - \log(Y^\top P^0)}{\epsilon} \\
			&= \lim_{\epsilon \to 0} \frac{\log(Y^\top P^0 + \epsilon\cdot Y^\top(\delta_i - P^0))) - \log(Y^\top P^0)}{\epsilon} \\
			&= \frac{Y^\top (\delta_i - P^0)}{Y^\top P^0} = \frac{y_i - \bar{y}}{\bar{y}}
		\end{align*}
	\end{itemize}

	\paragraph{In the response space}
	
	Note that in the response space (by transformation with $g^{-1}$) we can use the delta-method to get the corresponding variance estimates. Using the delta method is valid under decreasing $\mbox{var}(\hat{f})$. Thus for a testpoint $x$ if $\hat{f}(x)$ is the prediction and $\hat{V}(x)$ is the IJ variance estimate in the link space [\eqref{eq:varestcorr}] then in the response space the variance estimate will be $\hat{V}(x) \cdot \big((g^{-1})'(\hat{f}(x))\big)^2$.
	
	\subsection{Some theoretical exploration} \label{sec:gbftheory}
	
	We will now briefly explore some theoretical properties of our estimator and the variance estimate. 
	
	\paragraph{A Regularity Condition}
	
	First we need a regularity condition very similar to Condition 1 in \cite{ghosal2020boosting}. Note that the residuals $\ell_i'(\hat{\eta}_i^{(0)})/\big(-\ell_i''(\hat{\eta}_i^{(0)})\big)$ is a function of $\hat{\eta}_i^{(0)} = \hat{\eta}_{MLE}^{(0)}$ and thus depends on the whole dataset. Thus each of the trees in the random forest $\hat{f}^{(1)}$ depends on the whole dataset as well and so it's not a valid U-statistic kernel. This makes $\hat{f}^{(1)}$ not a valid U-statistic. A similar argument holds for $\hat{f}^{(2)}$. However, if we instead defined the random forests based on ``noise free'' residuals such as $\ell_i'(\EE[\hat{\eta}_i^{(0)}])/\big(-\ell_i''(\EE[\hat{\eta}_i^{(0)}])\big)$ then the base random forest would be a U-statistic (to make the second (boost) random forest a U-statistic as well we need a slightly more modified version of $\eta_i^{(1)}$). We should first make the following general assumption showing asymptotic equivalence of these two types of random forests.
	
	\begin{cond} \label{cond:regularity}
		For each of $j$ = $1,2$, define $\hat{f}^{(j)}$ to be the predictor obtained when using the residuals $\ell'_i(\hat{\eta}_i^{(j-1)})/\big(-\ell_i''(\hat{\eta}_i^{(j-1)})\big)$ and weights $-\ell_i''(\hat{\eta}_i^{(j-1)})$. We assume there are fixed functions $g^{(j-1)}$ such that for $\check{\eta}_i^{(j-1)} = g^{(j-1)}(x_i)$, the forest $\check{f}^{(j)}(x)$ obtained when using residuals $\ell'_i(\check{\eta}^{(j-1)})/\big(-\ell_i''(\check{\eta}_i^{(j-1)})\big)$ and weights $-\ell_i''(\check{\eta}_i^{(j-1)})$ follows
		$$
		\frac{\hat{f}^{(j)}(x) - \check{f}^{(j)}(x)}{\sqrt{var(\hat{f}^{(j)}(x))}}  \xrightarrow{p} 0.
		$$     	
	\end{cond}
	
	This condition is slightly more general than we need. A natural choice for the functions $g^{(j-1)}$ could be $g^{(0)}(x) = \EE \left[\hat{\eta}_{MLE}^{(0)}\right]$ and $g^{(1)}(x) = \EE \left[\hat{\eta}_{MLE}^{(0)} + \hat{f}^{(1)}(x)\right]$. However, $g^{(1)}(x) = E\left[\hat{\eta}_{MLE}^{(0)} + \check{f}^{(1)}(x)\right]$ could also be used so that $\check{\eta}_{MLE}^{(0)} + \check{f}^{(1)}(x) + \check{f}^{(2)}(x)$ represents a boosting analogue that is exactly a sum of $U$-statistics. Our main requirement is that the generalised boosted forest be approximated to within its standard error by functions that represent a sum of $U$ statistics.
	
	In the rest of this section we will state and prove our theoretical results for the estimator $\check{f}(x) = \hat{\eta}_{MLE}^{(0)} + \check{f}^{(1)}(x) + \check{f}^{(2)}(x)$. A discussion regarding this condition can be found in $\S 3$ of \cite{ghosal2020boosting}.
	
	\paragraph{Asymptotic Normality and Variance Consistency}
	
	Suppose the random forests $\hat{f}^{(1)}$ and $\hat{f}^{(2)}$ both have $B_n$ trees and each tree $T(x; Z_I)$ is constructed with a subsample $I$ of size $k_n$ (without replacement). Define the variance of this tree kernel to be $\zeta_{k_n,k_n}$ and we also know that the variance of the complete U-statistic with $T$ as the kernel is given by $\frac{k_n^2}{n} \zeta_{1,k_n}$ [\cite{hoeffding1948class, lee1990u}]. Without loss of generality we can also assume $\EE T(x; Z_1,\dots,Z_{k_n}) = 0$.
	
	Finally we assume the following Lindeberg-Feller type condition [due to \cite{mentch2016quantifying}]
	\begin{cond}\label{cond:mainLF}
		Suppose that the dataset $Z_1,Z_2,\dots \overset{iid}{\sim} D_Z$ and let $T(Z_1,\dots,Z_{k_n})$ be the tree kernel of a random forest based on a subsample of size $k_n$. Define $T_{1,k_n}(z) = \EE T(z,Z_2,\dots,Z_{k_n})$. We assume that for all $\delta > 0$
		$$
		\lim_{n \to \infty} \frac1{\zeta_{1,k_n}} \EE\left[ T^2_{1,k_n}(Z_1) \ind\{|T_{1,k_n}(Z_1)| > \delta\zeta_{1,k_n}\} \right] = 0
		$$ 
	\end{cond}
	
	Then under the conditions \ref{cond:regularity} and \ref{cond:mainLF} we can prove the following result	
	\begin{thm}\label{thm:main}
		Assume that the dataset $Z_1,Z_2,\dots \overset{iid}{\sim} D_Z$ and that $\EE T^4(x; Z_1,\dots,Z_{k_n}) \leq C < \infty$ for all $x, n$ and some constant $C$. If $k_n, B_n \to \infty$ such that $\frac{k_n}{n} \to 0$ and $\frac{n}{B_n} \to 0$ as $n \to \infty$ as well as $\lim\limits_{n \to \infty} \frac{k_n \zeta_{1,k_n}}{\zeta_{k_n,k_n}} \neq 0$ then we have
		$$
		\frac{\hat{f}(x)}{\sigma_n(x)} \overset{\DDD}{\to} \NNN(0,1)
		$$
		for some sequence $\sigma_n(x)$. Further $\sigma_n^2(x)$ is consistently estimated by $\hat{V}(x)$ as given by \eqref{eq:varest}.
	\end{thm}
	
	Our assumptions and conditions for this theorem is almost exactly the same as Theorem 1 in \cite{ghosal2020boosting} and so we will refer to that paper for the proof of asymptotic normality. The proof for the variance consistency part can be found in \cref{sec:varcons}.
	
	Also note that the above result is for the link space estimate and variance estimate. But we can easily observe that $\sigma_n(x) \to 0$ and so we can use an appropriate second-order Taylor expansion and a delta-method type argument to prove asymptotic normality of the response space estimate (mentioned at the end of \cref{sec:gbfdefine}) and consistency of the response space variance estimate (mentioned at the end of \cref{sec:varest}).
	
	\newpage
	\subsection{Algorithm}	
	
	\begin{algorithm}[ht] 
		\caption{Generalised Boosted Forest (\textit{and its variance estimate})}
		\SetKwInOut{Input}{Input}
		\SetKwInOut{Output}{Output}
		\SetKwInOut{Obtain}{Obtain}
		\SetKwInOut{Calculate}{Calculate}
		
		\Input{The data $\big(Z_i = (Y_i, X_i)\big)_{i=1}^n$, the link function $g^{-1}$, the log-likelihood function $\ell$, the tree function $T$, the number of trees in both forests $B$, the subsample size for each tree $k$, and the test point $x$.}
		
		\Obtain{The MLE-type constant $\hat{\eta}_{MLE}^{(0)} = \argmax_{t} \sum_{i=1}^n \ell(t; Z_i)$. This may have a closed-form or may be done numerically.}
		\Calculate{The directional derivatives of the MLE-type constant $U_i^{(0)}$.}
		
		Calculate weighted residuals $r_i^{(1)} = \frac{\ell_i'(\hat{\eta}_{MLE}^{(0)})}{-\ell_i''(\hat{\eta}_{MLE}^{(0)})}$ and training weights $w_i^{(1)} = -\ell_i''(\hat{\eta}_{MLE}^{(0)})$
		
		\For{$b = 1$ \KwTo $B$}{
			Choose a subset of size $k$ from $[n]$ such that the probability of the $i$\textsuperscript{th} datapoint being included is proportional to $w_i^{(1)}$.
			
			\Calculate{$T^{(1)}_b$ as the tree with the training data $r_I^{(1)}$ and $N_{i,b}^{(1)} = \ind\{i \in I\}$.}
		}
		\Obtain{The first (base) random forest $\hat{f}^{(1)}(x) = \frac1B \sum_{b=1}^B T^{(1)}_b(x)$.}
		\Calculate{The directional derivatives of $\hat{f}^{(1)}(x)$ as $U_i^{(1)}(x) = n\cdot cov_b(N_{i,b}^{(1)}, T^{(1)}_b(x))$.}
		
		Calculate weighted residuals $r_i^{(2)} = \frac{\ell_i'(\hat{\eta}_{MLE}^{(0)} + \hat{\eta}_i^{(1)})}{-\ell_i''(\hat{\eta}_{MLE}^{(0)} + \hat{\eta}_i^{(1)})}$ and training weights $w_i^{(1)} = -\ell_i''(\hat{\eta}_{MLE}^{(0)} + \hat{\eta}_i^{(1)})$, where $\hat{\eta}_i^{(1)} = \hat{f}^{(1)}(X_i)$.
		
		\For{$b = 1$ \KwTo $B$}{
			Choose a subset of size $k$ from $[n]$ such that the probability of the $i$\textsuperscript{th} datapoint being included is proportional to $w_i^{(2)}$.
			
			\Calculate{$T^{(2)}_b$ as the tree with the training data $r_I^{(2)}$ and $N_{i,b}^{(2)} = \ind\{i \in I\}$.}
		}
		\Obtain{The second (boosted) random forest $\hat{f}^{(2)}(x) = \frac1B \sum_{b=1}^B T^{(2)}_b(x)$.}
		\Calculate{The directional derivatives of $\hat{f}^{(2)}(x)$ as $U_i^{(2)}(x) = n\cdot cov_b(N_{i,b}^{(2)}, T^{(2)}_b(x))$.}
		\Output{The \textit{generalised boosted forest} estimate in the \textbf{link} space at the test point $x$ given by $\hat{f}(x) = \hat{\eta}_{MLE}^{(0)} + \hat{f}^{(1)}(x) + \hat{f}^{(2)}(x)$ and its (bias-corrected) variance estimate given by $\hat{V}(x) = \frac1{n^2} \sum_{i=1}^n \left( U_i^{(0)} + U_i^{(1)} + U_i^{(2)} \right)^2 + \frac1B \cdot \left(1 - \frac{n}{k}\right) \cdot \big(var_b(T_b^{(1)}(x)) + var_b(T_b^{(2)}(x)) \big)$.}
		\Output{The \textit{generalised boosted forest} estimate in the \textbf{response} space at the test point $x$ given by $g^{-1}(\hat{f}(x))$ and its variance estimate given by $\hat{V}(x) \cdot \big( g^{-1}(\hat{f}(x)) \big)^2$.}
	\end{algorithm}

	\section{Empirical Studies for Generalised Boosted Forest} \label{sec:gbfresults}
	
	We use the random forest implementation in the \verb|ranger| package in \verb|R| [\cite{rangerRpackage}]. In this package the weights (the second derivatives of the likelihood function) are used as probabilities for the corresponding datapoints when selected in the training set of a particular tree.
	
	\subsection{Performance on Simulated Datasets}\label{sec:simresults}
	
	We use a signal given by $f(X) = \sum_{i=1}^5 X_i$, where $X \sim U([-1,1]^{15})$. 
	
	Our simulation runs for 200 replicates - in each of them we generate a dataset of size $n = 1000$ and train a generalised boosted forest with varying levels of subsample sizes (20\%, 40\%, 60\% or 80\% of the size of the training data) and $B = 1000$ trees for each of the two random forests. For each replication we also use a separate fixed set of 100 test points from $U([-1,1]^{15})$ and then append 5 more fixed points to them, which are given by
	$$
	p_1 = \mathbf{0}_{15},\; p_2 = \left(\frac13, \mathbf{0}_{14}^\top \right)^\top,\; p_3 = \frac1{3\sqrt{15}}*\mathbf{1}_{15},\; p_4 = 2p_3,\; p_5 = 3 p_3
	$$
	Experiments in \cite{ghosal2020boosting} indicate that the signal to noise ratio affects the performance of boosted forests (referred to as the Gaussian case above) since if the noise in the dataset increase so does that in the estimate and the benefit from boosting decreases. The same is true for the generalised boosted forest. We have therefore experimented with different combinations of signal-to-noise ratios as given below.	
	\begin{itemize}
		\item \textbf{Binomial family:} One way to improve signal was to have the success probabilities far away from each other (since the noise is given by $p(1-p)$ and is minimised for $p=0,1$). For this purpose we multiplied the signal $f(X)$ by a scaling factor $s \in \{1,2,4,8,16\}$ before applying the logistic function to transform into probabilities. Thus the probabilities are pushed towards 0 and 1 as the scale increases. 
		
		We can also achieve higher signal by having a higher number of trials ($n_i$) - for this purpose we generated $n_i$ uniformly from the set $[M] = \{1,2,\dots,M\}$ where $M$ was varied to be in the set $\{1,2,4,8,16\}$. Once we have the values of $n_i$, we generate $Y_i \sim Binomial(n_i, p_i)$ where $p_i = g^{-1}(s \cdot f(X_i))$ for an appropriate scale $s$, and $g^{-1}$ being the logistic function.
		
		\item \textbf{Poisson family:} If $Y_i \sim Poisson(\lambda_i)$ then $var(\EE(Y_i))/mean(var(Y_i)) = var(\lambda_i)/mean(\lambda_i)$ is the signal-to-noise ratio. So if we multiply all $\lambda_i$ values with a scaling factor then the signal-to-noise ratio increases. Thus we set $Y_i \sim Poisson(\lambda_i)$, where $\lambda_i = g^{-1}(f(X_i) + \log(s)) = s \cdot g^{-1}(f(X_i))$, where $s \in \{1,2,4,8,16\}$ and $g^{-1}$ is the exponential function.
	\end{itemize}
	
	For each of these settings we obtain predictions and a variance estimates in the link space, which we can also transform into their corresponding values in the response space. We use the following metrics to measure our performance
	
	\begin{enumerate}
		\item The improvement in test-set log-likelihood from the MLE-type stage to fitting the base forest, and also from fitting the base forest to the boosted stage. In the leftmost panel in Figure \ref{fig:testsetplots} we show the binomial case only for the value of $M=4$ since all the other number of trials have a very similar shape to the plot (when suitably scaled). Note that the y-axis of the plots are in the pseudo log scale (inverse hyperbolic sin)
		
		\item We can define MSE in two ways - by $\frac1n \sum (\hat{f}(X_i) - f(X_i))^2$ in the link space and by $\frac1n \sum (g^{-1}(\hat{f}(X_i)) - g^{-1}(f(X_i)))^2$ in the response space. Consider the three different estimators $\hat{f}$ corresponding to the MLE-type, base random forest and final boosted estimator and the improvement in MSE that can be calculated between these stages in the same way as for log-likelihood above. In the second and third diagrams in Figure \ref{fig:testsetplots} we've also restricted the binomial family to the case $M=4$ (the other values of $M$ show plots with similar shape and scale). Here too the y-axis of the plots are in the pseudo log scale (inverse hyperbolic sin). 
		
		Note that each point in these plots are a summary of 200 numbers and for the Poisson family with high subsample fractions and high scales an unusual situation may arise. In this case the response space MSE may \textit{increase} between the base and first boosting stage - although this doesn't always happen, the amount of increment has a very heavy right tail, and the base to boost stage almost always corrects for this increase in MSE. For this reason, we show the median MSE improvement in this plot rather than the mean. Note that this is in spite of the link space showing \textit{decreases} in MSE between the different stages for all cases - this is be because the response space estimate is an exponentiation of the link space estimate and has too much variability, i.e, outliers play a larger role in skewing MSE. We have included a more detailed explanation for this phenomenon in \cref{sec:poissonMSE}. 
		
		\begin{figure}[ht]
			\centering
			\caption{Improvements in loglikelihood and MSE (link and response spaces) in the pseudo-log scale}
			\label{fig:testsetplots}
			\includegraphics[width=\columnwidth]{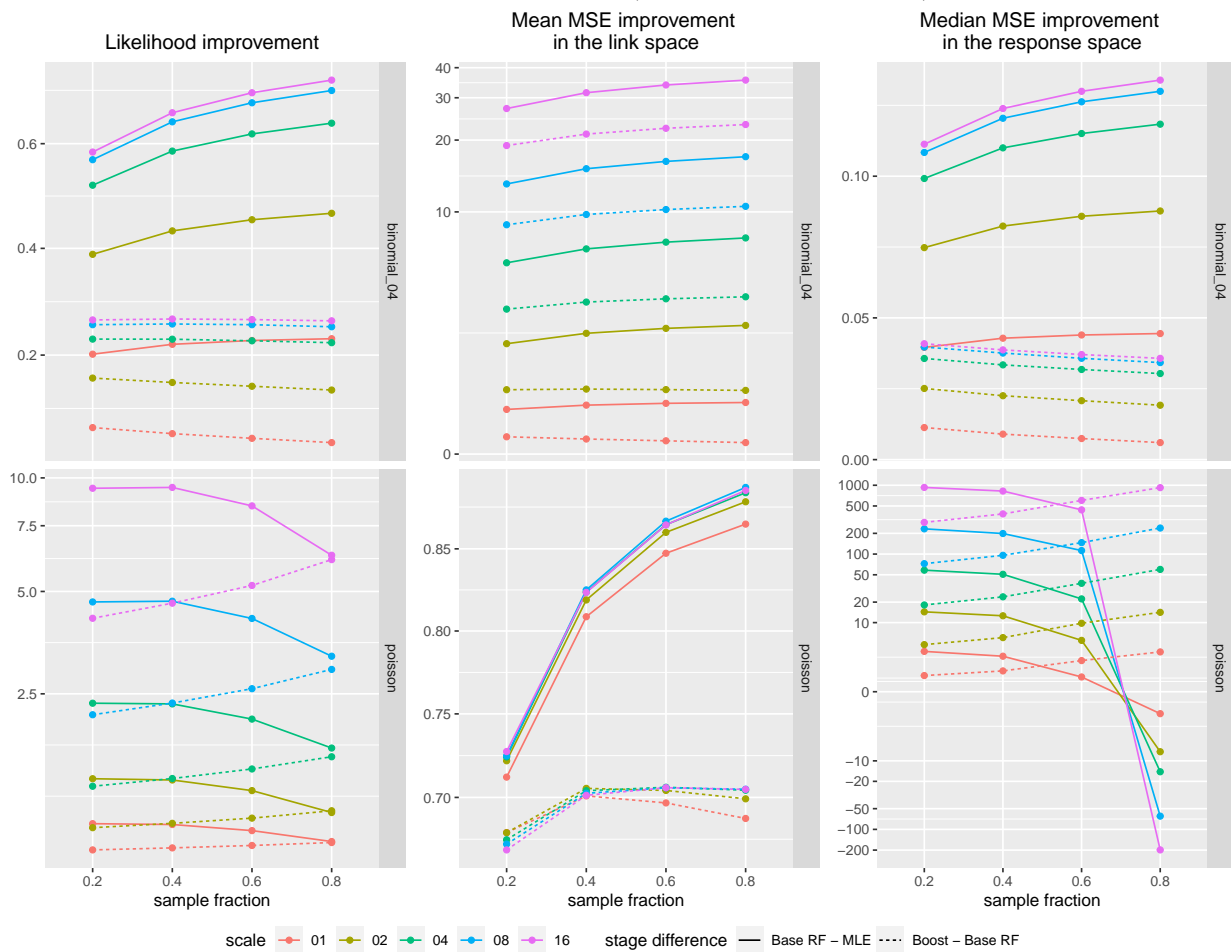}
		\end{figure}
		
		\item Confidence interval coverage as the average number of times (out of the 200 replicates) the real signal is contain within the 95\% confidence interval. We plot the coverage as a function of the true value of the signal (shown in the pseudo-log scale) in Figure \ref{fig:testsetcoverage}. We restrict ourselves to the link space, a few representative families and scales, and a sample fraction of 0.4 (since we've observed that coverage doesn't really depend on the sample fraction). 
		
		We see that the coverage always improves with the second (boosted) random forest. For the binomial family there seems to be a threshold beyond which the confidence interval doesn't cover the true signal at all and this threshold is independent of both ways of trying to improve the signal-to-noise ratio. A similar phenomenon is apparent for the Poisson family although here the threshold depends on the scale. Possible reasons for this behaviour are discussed in \cref{sec:gbflimits}. 
		
		\begin{figure}[ht]
			\centering
			\caption{Coverage of 95\% confidence intervals vs the true signal}
			\label{fig:testsetcoverage}
			\includegraphics[width=\columnwidth]{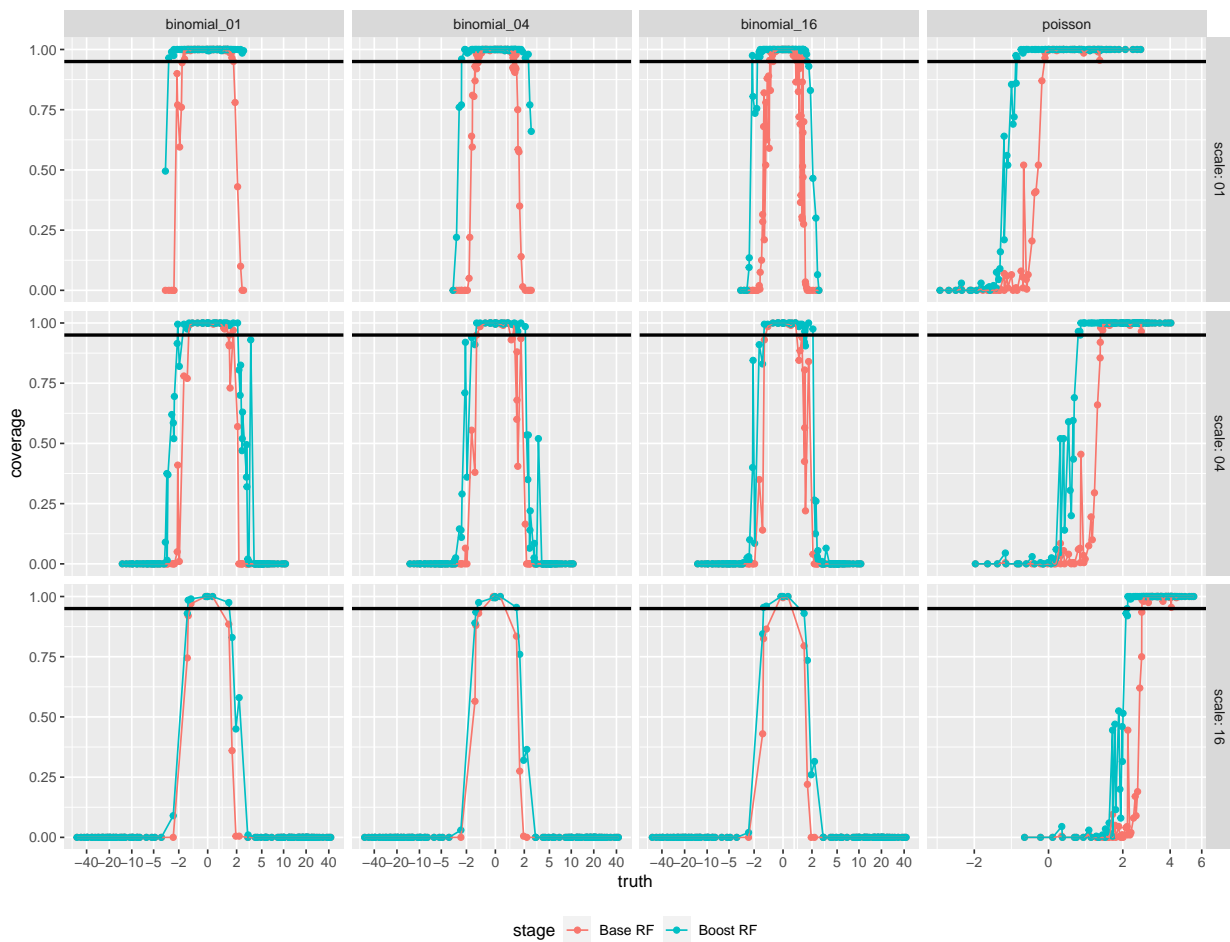}
		\end{figure}
	\end{enumerate}
	
	More detailed plots and discussion using further metrics can be found in \cref{sec:detailedlinearplots}.
	
	For the 5 fixed points $p_1,\dots,p_5$ we can also find improvements in expected values of log-likelihood and MSE but the effects won't be very pronounced for individual points. Instead we will look at the following metrics 
	\begin{enumerate}
		\item Absolute bias averaged over the 200 replicates. In Figure \ref{fig:5ptsbias} we again focus only on $M=4$ for the binomial family (the other values of $M$ show plots with similar shape and scale). Note that for similar reasons as MSE above, the average absolute bias can be calculated in the link (top two) and response (bottom two) spaces. 
		
		We can see that in the link space and for the binomial family (top left of Figure \ref{fig:5ptsbias}), the absolute bias doesn't depend on the subsample fraction but increases with scale and the distance of the test point from the origin. For the other three cases relationship with scale and distance from origin is similar although increasing subsample fractions generally decreases the absolute bias. Most importantly in all cases we can see that boosting reduces absolute bias.
		
		\begin{figure}[p]
			\centering
			\caption{Absolute bias for the 5 fixed points - link space on top, response space on bottom}
			\label{fig:5ptsbias}
			\includegraphics[width=\columnwidth]{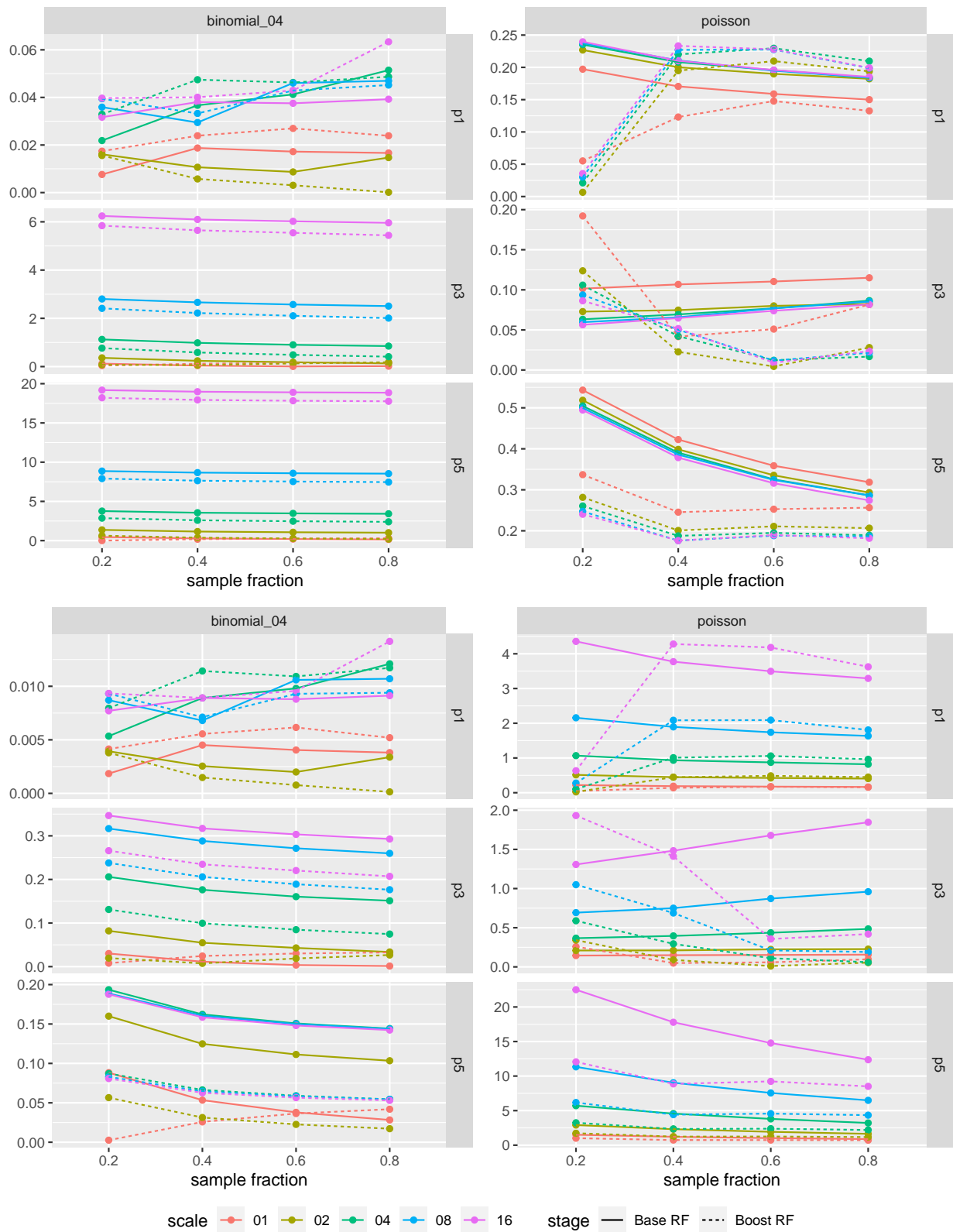}
		\end{figure}
		
		\item Consistency of the variance estimate (bias corrected) calculated by taking the ratio of \textit{average of the variance estimates} over the \textit{variance of the estimates} (over the 200 replicates). In Figure \ref{fig:5ptsvarcons} we restrict the binomial family to $M=4$ since the other values of $M$ have plots with similar shape and scale. This plot is only for the link space although the plot for the response space is very similar. 
		
		Here we see that the variance is more consistent for higher subsample fraction. For the binomial family variance consistency is slightly better for points nearer to the origin although the opposite effect is observed for the Poisson family. The effect of scale on variance consistency is difficult to ascertain. Lastly the effect of the boosting step on this metric seems to depend on the test-point and the family - for test-points farther from the origin this metric becomes better with boosting for the binomial, but worse for the Poisson family.
		
		\begin{figure}[ht]
			\centering
			\caption{Variance consistency for the 5 fixed points (link space)}
			\label{fig:5ptsvarcons}
			\includegraphics[width=\columnwidth]{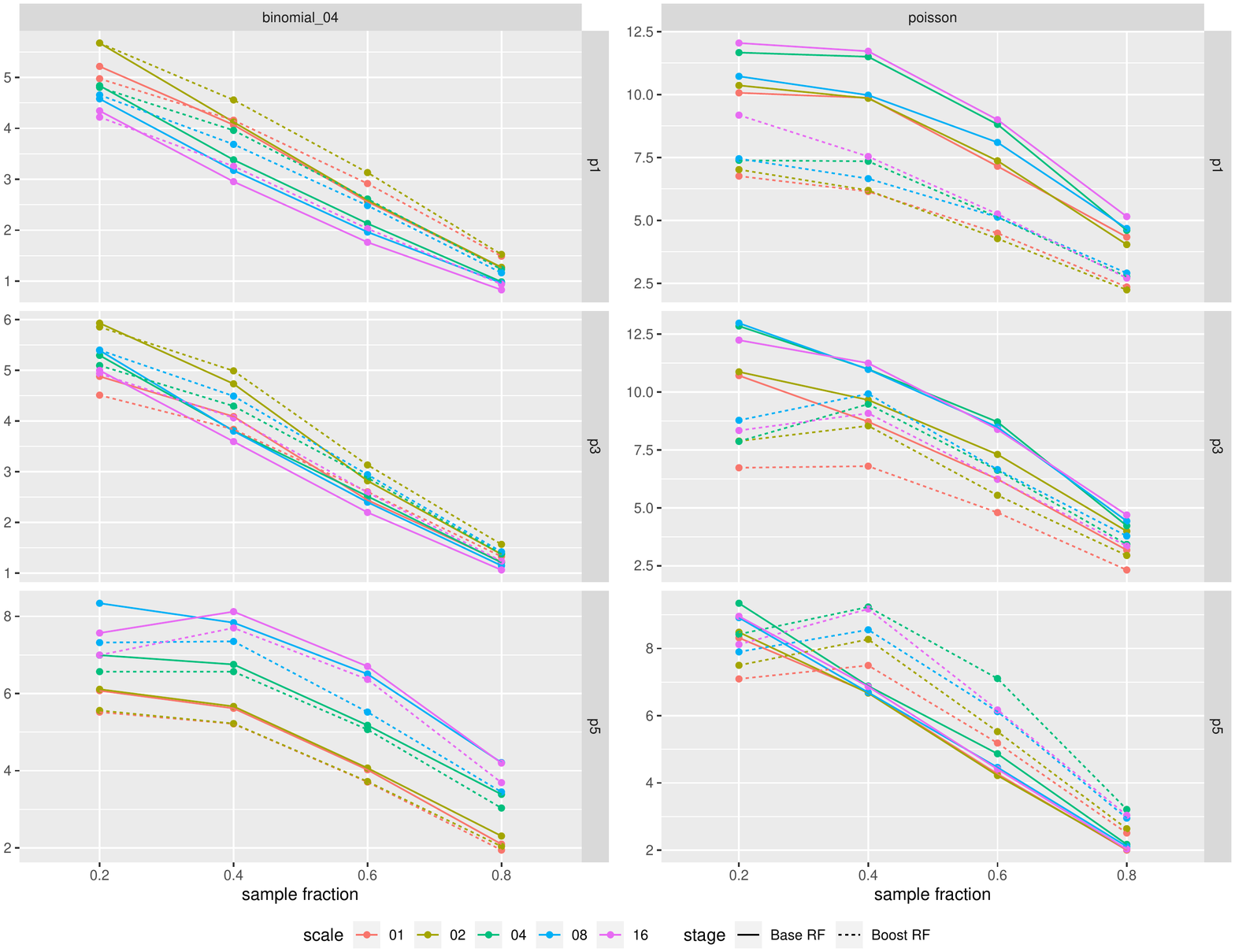}
		\end{figure}
		\item Kolmogorov-Smirnov statistic for normality. This can also be calculated for the link and response spaces (i.e., before and after the delta transformation). The values for the K-S statistic in the link space fall in the range of 0.1-0.3 - it doesn't change too much for the different points, but it increases slightly with subsample fraction and scales $s$, it is also slightly lower for the boosted stage than the base forest stage. A detailed plot for this metric can be found in \cref{sec:detailedlinearplots}.
	\end{enumerate}
	
	See \cref{sec:detailedlinearplots} for more detailed plots and discussions with other performance metrics.
	
	\subsection{Performance on Real Datasets}\label{sec:realresults}
	
	We present below results for some real datasets from the UCI database. Given the data $(Y_i, X_i)_{i=1}^n$ we employed a 10-fold cross-validation to get predictions $\hat{f}(X_i)$ and variance estimates $\hat{V}_i$ in the link space for all three stages. For performance metrics we use the log-likelihood (defined similar to above) MSE, average variance and prediction intervals (in the response space). The last three are given by
	$$
	\text{MSE} = \frac1n \sum_{i=1}^n (Y_i - g^{-1}(\hat{f}(X_i)))^2, \; \text{Avg Var} = \frac1n \sum_{i=1}^n \hat{V}_i^{(R)} = \frac1n \sum_{i=1}^n \hat{V}_i \cdot \big(g^{-1}(\hat{f}(X_i))\big)^2,
	$$
	$$
	95\% \text{ prediction interval for } Y_i, \text{i.e.}, P_i = \left( g^{-1}(\hat{f}(X_i)) \pm \Phi^{-1}(0.975) \sqrt{\hat{V}_i^{(R)} + \hat{V}_e^{(R)}} \right),
	$$
	where $g^{-1}$ is the appropriate link function and $V_e^{(R)}$ is the error variance (in the response space) estimated by $\hat{V}_e^{(R)} =$ MSE. Once we have the prediction intervals we calculate the prediction coverage (PC) to be $\sum_{i=1}^n \ind_{\{Y_i \in P_i\}}$.
	
	We've used the \textsf{Abalone} and \textsf{Solar flare} datasets to fit the Poisson family, with results reported in Table \ref{tab:poisson}.
	\begin{table}[ht]
		\caption{Performance statistics for the \textsf{Abalone} and \textsf{Solar Flare} datasets.}
		\label{tab:poisson}
		\begin{center}
			\begin{tabular}{lrrrr}
				\toprule
				\textbf{\textsf{Abalone}} & MSE & Avg Var & PC & LL\\
				\midrule
				stage0 & 10.413 & 0.003 & 0.9494 & 12.8729\\
				stage1 & 4.688 & 3.583 & 0.9825 & 13.1637\\
				stage2 & 4.549 & 5.848 & 0.9882 & 13.1745\\
				\bottomrule
			\end{tabular}
			\quad
			\begin{tabular}{lrrrr}
				\toprule
				\textbf{\textsf{Solar flare}} & MSE & Avg Var & PC & LL\\
				\midrule
				stage0 & 0.703 & 0.001 & 0.9340 & -0.6659\\
				stage1 & 5.195 & 20.312 & 0.9972 & -0.8382\\
				stage2 & 1.768 & 6.661 & 0.9849 & -0.6526\\
				\bottomrule
			\end{tabular}
		\end{center}
	\end{table}
	We can see that the boosting stages improve all the measures for the \textbf{Abalone} data (except for the average variance). But in the \textbf{Solar flare} data the response takes the value 0 a lot of the time and only occasionally has values between 1 and 8 - that is why the first stage random forest fails to improve log-likelihood but further boosting helps in this case.

	We've also used the \textsf{Spam} dataset for the binomial (Bernoulli) family where our results are given in Table \ref{tab:binom}.
	\begin{table}[ht]
		\caption{Performance statistics for the \textsf{Spam} dataset.}
		\label{tab:binom}
		\begin{center}
			\begin{tabular}{lrrrr}
				\toprule
				\textbf{\textsf{Spam}} & MSE & Avg Var & PC & LL\\
				\midrule
				stage0 & 0.239 & 0.00006 & 1.0000 & -0.6709\\
				stage1 & 0.059 & 0.02725 & 0.9813 & -0.2484\\
				stage2 & 0.040 & 0.03162 & 0.9837 & -0.1648\\
				\bottomrule
			\end{tabular}
		\end{center}
	\end{table}
	Here boosting improves both MSE and the log-likelihood. It also makes the coverage for a 95\% prediction interval less conservative. Further note that the log-likelihood using the probability predictions from the \verb|ranger| package in \verb|R| is found to be -0.1957 which is less that the final log-likelihood given by our method, i.e, our method fits the dataset better than traditional classification forests in the log-likelihood metric.
	
	\section{Conclusion}
	
	We demonstrated a new method, \textit{Generalised Boosted Forest}, to fit responses which can be modelled by an exponential family. With the goal of maximising the training set log-likelihood we start with an MLE-type estimator in the link space (as opposed to the actual MLE which is in the response space). Then with a second order Taylor series approximation to the log-likelihood we showed that a random forest, fitted with pseudo-residuals and appropriate weights, can improve this log-likelihood. Then, similar to \cite{ghosal2020boosting}, we add another random forest as a boosting step which further improves the log-likelihood. Our method also uses the Infinitesimal Jackknife to get variance estimates, at no additional computational cost.
	
	Through simulations and real data examples we've shown the effectiveness of our algorithm. The random forests always improve the test-set log-likelihood in all of the cases we've tested. We also see noticeable improvement in test-set MSE. Our estimates for uncertainty quantification shows conservative coverage of the resulting confidence/prediction intervals and the near-Gaussianity of the distribution of predictions. 
	
	Although outside the scope of this paper, the method could be improved in many ways. We could of course try more boosting steps; the boosting steps could also have a multiplier determined either by hyper-parameter tuning or with Newton-Raphson. Also note that we have not used over-dispersed models for variance, but obtaining an overdispersion parameter may be important in providing appropriate confidence intervals in some cases.  We may also wish to consider alternative base-stage models such as a generalised linear model rather than a straightforward constant. In this case, performance improvement from boosting this model with random forests can also be thought of as a goodness of fit test for the base model.

	\section*{Acknowledgements}
	
	We are indebted to \href{mailto:brb225@cornell.edu}{Benjamin Baer} for the idea of Newton boosting updates.
	
	\printbibliography

	\afterpage{\clearpage}
	\newpage
	
	\appendix
	
	\section{Consistency of Infinitesimal Jackknife}\label{sec:varcons}
	
	As stated in Theorem \ref{thm:main} we will prove consistency of the variance estimate in \eqref{eq:varest}.
	
	\textit{Proof:} Suppose the MLE-type estimator in the signal space is given by $f(\bar{y})$ where $f$ has a continuous derivative. By the classical central limit theorem $\bar{y} \sim N(\mu, \frac{\sigma^2}{n})$ and so by the delta method $f(\bar{y}) \sim N(f(\mu), \frac{\sigma^2}{n} \cdot (f'(\mu))^2)$ approximately. Also it is easily seen that the IJ directional derivatives for $f(\bar{y})$ are $(U_i)_{i=1}^n$ where $U_i = f'(\bar{y})(y_i - \bar{y})$. So the IJ estimate for the variance of $f(\bar{y})$ is
	$$
	\frac1{n^2} \sum_{i=1}^n U_i^2 = \frac1{n^2} \sum_{i=1}^n (f'(\bar{y}))^2 \cdot (y_i - \bar{y})^2 = \frac{s^2}{n} \cdot (f'(\bar{y}))^2
	$$
	Since $\bar{y}$ and $s^2$ are asymptotically independent hence $\EE\left[ \frac{s^2}{n} \cdot (f'(\bar{y}))^2 \right] = \frac{\sigma^2}{n} \cdot (f'(\mu))^2$.
	
	Now note that the IJ variance estimate for the two random forests and also the IJ covariance estimate between them has been shown to be consistent in \cite{ghosal2020boosting}. We only need to show consistency of the IJ covariance estimate between the MLE-type estimator $f(\bar{y})$ and the random forests. The following procedure shows this for the base random forest but it will be a similar process for the boosting step as well - note that by Condition \ref{cond:regularity} the noise from $f(\bar{y})$ doesn't affect the base random forest or it's directional derivatives.
	
	We know from \cite{ghosal2020boosting} that the IJ directional derivatives for a random forest is given by $U_i' = n \cdot cov_b (N_{i,b}, T_b(x))$ where $N_{i,b}$ is the number of times the $i$\textsuperscript{th} training point is included in the sample for the $b$\textsuperscript{th} tree and $T_b$ is the prediction from the $b$\textsuperscript{th} tree for the test point $x$. The IJ variance estimate $\hat{\sigma}_F^2 = \frac1{n^2} (U_i')^2$ is consistent for $\sigma_F^2$, the theoretical variance of the first Hajek projection of the random forest $\hat{F}(x)$, i.e, 
	\begin{align*}
		\sigma_F^2 &= \sum_{i=1}^n \left[ \EE_{U \sim D} (\hat{F}(x) \mid U_1 = Z_i) - \EE_{U \sim D} (\hat{F}(x)) \right]^2 \\
		&= \frac{k^2}{n^2} \sum_{i=1}^n \left[ \EE_{U \sim D} (T(x) \mid U_1 = Z_i) - \EE_{U \sim D} (T(x)) \right]^2,
	\end{align*}
	where $(Z_i)_{i=1}^n = (Y_i, X_i)_{i=1}^n$ is the training data, $T$ is the tree kernel function for the U-statistic $\hat{F}$, $U$ is the training set (of size $k$) of a tree $T$, and $D$ is the distribution from which the training data is drawn. So we actually define
	$$
	\sigma_F^2 = \frac{k^2}{n^2} \sum_{i=1}^n \left[ \EE_{U \sim \hat{D}} (T(x) \mid U_1 = Z_i) - \EE_{U \sim \hat{D}} (T(x)) \right]^2 = \frac{k^2}{n^2} \sum_{i=1}^n (A_i + R_i)^2 \text{ (say)},
	$$
	where $A_i = \EE_{U \sim \hat{D}} (\mathring{T}(x) \mid U_1 = Z_i) - \EE_{U \sim \hat{D}} (\mathring{T}(x))$ with $\mathring{T} = \sum_{i=1}^k T_1$ being the first Hajek projection of $T$. Then from \cite{wager2017estimation} $\EE\left[ \frac{k^2}{n^2} \sum_{i=1}^n A_i^2 \right] = \sigma_F^2$ asymptotically and $\frac1{\sigma_F^2} \frac{k^2}{n^2} \sum_{i=1}^n R_i^2 \xrightarrow{p} 0$.
	
	For the covariance we need to show that $\frac{k}{n^2} \sum_{i=1}^n A_i U_i$ is consistent for $cov(f(\bar{y}), \hat{F}(x))$. Now note that by Taylor expansion $f(\bar{y}) = f(\mu) + (\bar{y} - \mu) f'(y^*)$ where $|y^* - \mu| < |\bar{y} - \mu|$. But $\bar{y} \to \mu \implies y^* \to \mu \implies f'(y^*) \to f'(\mu)$ by the Strong Law of Large Numbers and by continuity of $f'$. Also from \cite{wager2017estimation} we know that the first Hajek projection of $\hat{F}(x)$ is $\EE[T] + \frac{k}{n} \sum_{i=1}^n T_1(Z_i)$. So
	\begin{align*}
		cov(f(\bar{y}), \hat{F}(x)) &\approx cov\left(f'(\mu) \left(\frac1n \sum_{i=1}^n y_i - \mu\right), \EE[T] + \frac{k}{n} \sum_{i=1}^n T_1(Z_i)\right) \\
		&= f'(\mu) \cdot cov\left( \frac1n \sum_{i=1}^n y_i, \frac{k}{n} \sum_{i=1}^n T_1(Z_i) \right) \\
		&= f'(\mu) \cdot \frac1n \cdot \frac{k}{n} \cdot n \cdot cov(y_n, T_1(Z_n)) \\
		&= \frac{k}{n} \cdot f'(\mu) \cdot cov(y_n, T_1(Z_n))
	\end{align*}
	
	As a sidenote we observe here that the term $k/n \to 0$ as per the assumptions in Theorem \ref{thm:main} and so we could actually ignore this covariance term (and it's estimate) entirely in the final variance estimate. But we will still go ahead and show consistency for the covariance estimate defined above.
	
	Suppose $U_i = B_i + S_i$ where $B_i = f'(\mu) (y_i - \mu)$. We also know from \cite{wager2017estimation} that $A_i = \frac{n-k}{n} \left[ T_1(Z_i) - \frac1{n-1} \sum_{j \neq i} T_1(Z_j) \right]$. Then
	\begin{align*}
		\EE[A_iB_i] &= \frac{n-k}{n} \cdot \EE(T_1(Z_i)) \cdot f'(\mu) (y_i - \mu) = f'(\mu) \cdot \frac{n-k}{n} \cdot cov(T_1(Z_i), y_i - \mu) \\
		\frac{k}{n^2} \sum_{i=1}^n \EE[A_iB_i] &= \frac{k}{n^2} \cdot n \cdot f'(\mu) \cdot \frac{n-k}{n} \cdot cov(T_1(Z_n), y_n) = \frac{n-k}{n} \cdot cov(f(\bar{y}), \hat{F}(x))
	\end{align*}
	Also note that $S_i = f'(\bar{y}) (y_i - \bar{y}) - f'(\mu)(y_i - \mu) = y_i(f'(\bar{y}) - f'(\mu)) - (\bar{y}f'(\bar{y}) - \mu f'(\mu)) = y_i p_n - q_n \text{ (say)}$. Now by the strong law of large numbers $\bar{y} \to \mu$, $\frac1n \sum_{i=1}^n y_i^2  = s^2 + \bar{y}^2 \to \sigma^2 + \mu^2$ and also by continuity $p_n, q_n \to 0$. Thus
	\begin{align*}
		\frac1{n^2} \sum_{i=1}^n S_i^2 &= \frac1{n^2} \sum_{i=1}^n (y_i^2 p_n^2 + q_n^2 - 2y_ip_nq_n) \\
		&= p_n^2 \cdot \frac1n \cdot \left(\frac1n \sum_{i=1}^n y_i^2\right) + \frac1n \cdot q_n^2 - 2 \bar{y} p_n q_n \to 0
	\end{align*}
	Hence by Cauchy-Schwartz $\frac{k}{n^2} \sum_{i=1}^n \EE[A_iS_i] \to 0$. Finally since we assume $k/n \to 0$ we have proved asymptotic consistency.
	
	\section{Limitations on the Range of Predictions} \label{sec:gbflimits}
	
	Our coverage plots demonstrate significant under-coverage at large and small values of the signal. These also correspond to positions in covariate space that are near the edges of the data distribution.  While much of the observed coverage attenuation may be explained by edge effects -- when most of data points influencing a prediction are close to the center of the distribution and hence have signals closer to zero -- we also identify a truncation effect associated with the use of Newton residuals that may also effect other boosting frameworks. 
	
	The training data for each of the two random forests are of the form $r_i = \frac{\ell_i'(t_i)}{-\ell_i''(t_i)}$. In the Gaussian case this can take any real value as $\ell_i'$ is linear, $\ell_i''$ is a constant and $t_i$ ranges over $\RR$. But in case of other exponential families this range may be limited. Note that output predictions from a random forest are convex combinations of the input training signals - so if training signals have a lower (and/or upper) bound then the predictions will also have the same lower (and/or upper) bound.
	
	\paragraph{In the binomial case}
	$$
	\ell_i'(t_i) = y_i - n_i \cdot \frac{e^{t_i}}{1+e^{t_i}} \implies \ell_i''(t_i) = - n_i \cdot \frac{e^{t_i}}{(1+e^{t_i})^2} \implies r_i = \frac{\ell_i'(t_i)}{-\ell_i''(t_i)} = \frac{y_i - n_ip_i}{n_ip_i(1-p_i)}
	$$
	where $p_i = \frac{e^{t_i}}{1+e^{t_i}} \in [0,1]$. The minimum and maximum values of this is achieved when $y_i$ is 0 and $n_i$ respectively. Consequently
	$$
	\min_i r_i = \min_i \frac{- n_ip_i}{n_ip_i(1-p_i)} = - \frac1{1 - \max_i p_i}; \qquad \max_i r_i = \max_i \frac{n_i - n_ip_i}{n_ip_i(1-p_i)} = \frac1{\min_i p_i}
	$$
	Now suppose $p_0 = \frac{e^{\hat{\eta}_{MLE}^{(0)}}}{1+e^{\hat{\eta}_{MLE}^{(0)}}}$ is the MLE corresponding to the first constant. Then the range of training data for the first (base) random forest $\hat{f}^{(1)}$ is $\left[ -\frac1{1 - p_0}, \frac1{p_0} \right]$. The range of predictions from $\hat{\eta}_{MLE}^{(0)} + \hat{f}^{(1)}$ will thus be $\left[ \log\left(\frac{p_0}{1-p_0}\right) - \frac1{1 - p_0}, \log\left(\frac{p_0}{1-p_0}\right) + \frac1{p_0} \right]$. Clearly this interval does not cover the whole real line. This is why in \cref{sec:simresults} there are some true signals for which the confidence interval coverage is 0\% - these signals are outside the range of predictions for the forest and the variance estimates are not high enough to make up for the difference. It is also clearly seen by the above calculation that this theoretical limit of predicted values do not depend on the scale or the number of trials.
	
\begin{figure}[ht]
		\centering
		\caption{Theoretical range of predictions (link space) for the binomial family}
		\label{fig:binomialrange}
		\includegraphics[width=\columnwidth]{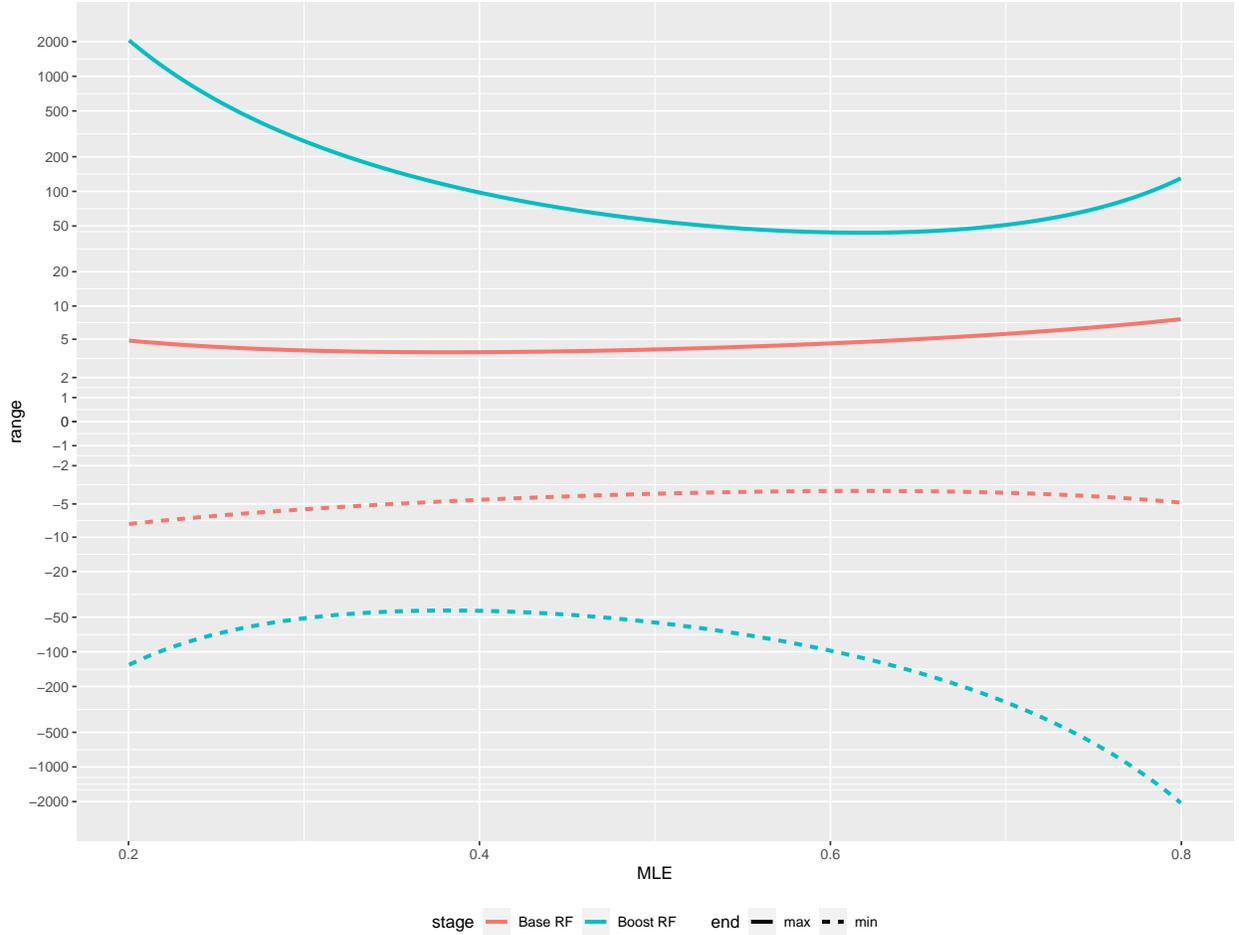}
	\end{figure}

	Note that the theoretical range of predictions by including the second (boost) random forest will be $$\left[ \log\left(\frac{p_0}{1-p_0}\right) - \frac1{1 - p_0} - \frac1{1 - \max_i p_i}, \log\left(\frac{p_0}{1-p_0}\right) + \frac1{p_0} + \frac1{\min_i p_i} \right],$$ where $\max_i p_i$ and $\min_i p_i$ will have complicated expressions in terms of $p_0$, but nevertheless it will still be a bounded interval. A plot of the these ranges as a function of $p_0$ is given in Figure \ref{fig:binomialrange} (y-axis in the pseudo-log scale), where we note a very wide range even at its smallest value. 
	
	\paragraph{In the Poisson case}
	
	$$
	\ell_i'(t_i) = y_i - e^{t_i} \implies \ell_i''(t_i) = - e^{t_i} \implies r_i = \frac{\ell_i'(t_i)}{-\ell_i''(t_i)} = \frac{y_i}{e^{t_i}} - 1
	$$
	Since $y_i$ is always non-negative, $r_i \geq -1$. Thus the possible range of predictions from $\hat{\eta}_{MLE}^{(0)} + \hat{f}^{(1)}$ is bounded below by $\hat{\eta}_{MLE}^{(0)} - 1$; and $\hat{\eta}_{MLE}^{(0)} - 2$ if we include the second random forest. This is why in \cref{sec:simresults} there are some true signals with value below $\hat{\eta}_{MLE}^{(0)} - 1$ for which the confidence interval coverage is 0\%. Note that $\hat{\eta}_{MLE}^{(0)}$ and hence the lower bound threshold depends on the scale (used for improving the signal-to-noise ratio). Although there isn't an upper bound and thus highly positive values of the signal obtains a high coverage by the confidence interval.
	
	\section{The effect of outliers in the Poisson case} \label{sec:poissonMSE}
	
	In our Poisson simulations we observe that the first random forest can sometimes perform worse than the initial constant in terms of response space MSE even when it improves MSE in the link space, and that this is then ameliorated by the second random forest, suggesting that the effect is not simply due to variance. This behaviour is also more pronounced at larger subsample fractions. Here we provide a heuristic explanation for this observation. 
	
	Recall that in the Poisson case the random forest is trained on responses $\left(\frac{y_i}{e^{t_i}} - 1\right)_{i=1}^n$ and with the $i$\textsuperscript{th} datapoint having the chance to be selected for a tree being $e^{t_i}$, where $t_i$ is the current prediction. After the MLE-type constant stage $t_i = \hat{\eta}_{MLE}^{(0)}$ is constant, i.e., all data points are equally weighted but we expect a small number of training points to have very high values. These, in turn, will exert significant influence on the structure of the trees for which they are in-bag: splits that isolate these outliers will exhibit greatest decrease in squared error. Test points nearby such an outlier will therefore receive very large predictions from each tree for which the outlier was in bag; something that will increase with bag fraction.  This is then exacerbated when the model is exponentiated to be in the response space. 
	
	Specifically suppose we have a very large $y$ in the training data. The initial pseudo-residual $z = y / e^{\hat{\eta}_{MLE}^{(0)}} - 1$ is also large, so that any tree for which the data point is in-bag will give a prediction $\alpha z + c$ where $\alpha$ accounts for the minimum leaf size and $c$ for the other pseudo-responses in the other leaves. A test point that is consistently in the same leaf as our outlier (when it is in-bag) will then give a prediction of $\alpha' z + c'$, where the multiplier $\alpha'$ now also accounts for the in-bag fraction; approximately proportional to the subsample fraction.
	
	By itself, this is not particularly detrimental. The error $\alpha' (y / e^{\hat{\eta}_{MLE}^{(0)}} - 1) + c' - f(x)$ may not be terribly large. However, when we exponentiate to get to the response space, the error $e^{\alpha' (y / e^{\hat{\eta}_{MLE}^{(0)}} - 1) + c'} - e^{f(x)}$ can dominate the remaining errors by virtue of exponentiating an already-large value of $y$. When we fit a second random forest (the boost step) then the pseudo-residual corresponding to the outlier is now much smaller, but there are also large negative pseudo-residuals associated with any training points with high in-leaf proximity to the outlier. This then allows the boosted step to lower the high MSE from the previous (base) step.
	
\begin{figure}[ht]
		\centering
		\caption{Detecting test-points causing high MSE in link or response spaces}
		\label{fig:poissonoutlier}
		\includegraphics[width=\columnwidth]{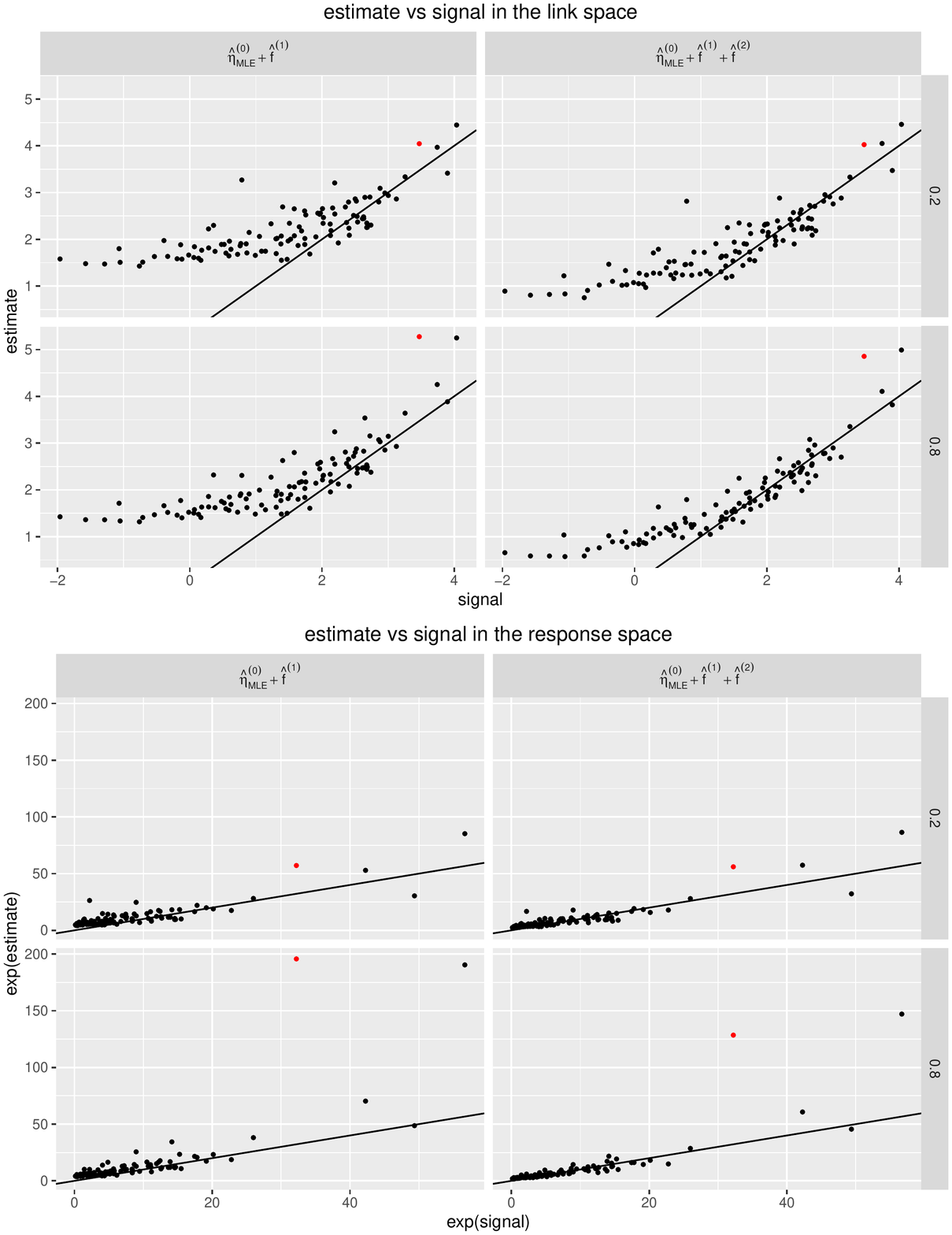}
	\end{figure}

	We demonstrate this situation empirically as follows. Our model is the same as \cref{sec:simresults} but for only one replicate, one scaling factor (4) and two subsample fractions. For a particular seed of the pseudo-random number generator in \texttt{R}, we obtain the performance statistics given in Table \ref{tab:example}.
	\begin{table}[ht]
		\caption{Performance statistics for an example Poisson simulation with poor first-stage performance.}
		\label{tab:example}
		\begin{center}
			\begin{tabular}{lrrrrrr}
				\toprule
				\multicolumn{1}{c}{ } & \multicolumn{2}{c}{Link space MSE} & \multicolumn{2}{c}{Response Space MSE} & \multicolumn{2}{c}{Log-likelihood} \\
				\cmidrule(l{3pt}r{3pt}){2-3} \cmidrule(l{3pt}r{3pt}){4-5} \cmidrule(l{3pt}r{3pt}){6-7}
				Subsample fraction & 0.2 & 0.8 & 0.2 & 0.8 & 0.2 & 0.8\\
				\midrule
				Stage 0 & 2.089 & 2.089 & 90.303 & 90.303 & 7.745 & 7.745\\
				Stage 1 & 1.381 & 1.205 & 44.115 & 454.862 & 10.163 & 9.038\\
				Stage 2 & 0.714 & 0.505 & 27.748 & 174.375 & 11.206 & 10.806\\
				\bottomrule
			\end{tabular}
		\end{center}
	\end{table}
	The MSE is always lowered as we go through more stages except for a high subsample fraction in the response space where MSE first increases substantially and then decreases somewhat subsequently but still not enough to have a better performance that just the constant. Notably the log-likelihood, which is the main target of our optimiser, always increases. For the data corresponding to the table above we've plotted the estimated vs true signals for the test-points in Figure \ref{fig:poissonoutlier}.	
	\afterpage{\clearpage}
	
\begin{figure}[ht]
		\centering
		\caption{The anomalous test point has higher proximity with outlier training points}
		\label{fig:poissonproximity}
		\includegraphics[width=\columnwidth]{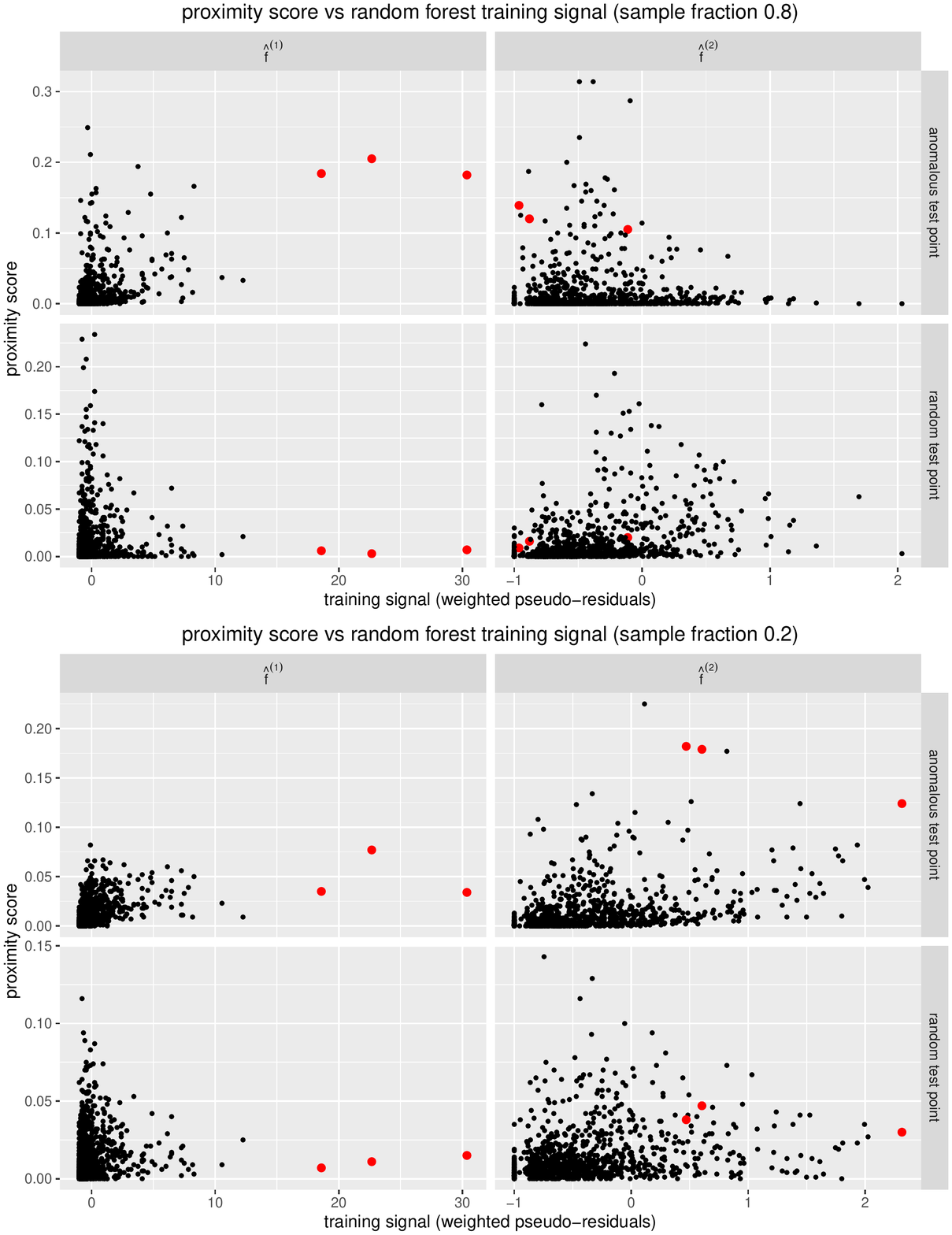}
	\end{figure}

	Clearly there is a test-point whose estimated prediction is very far away from the true value in the response space, for a high subsample fraction. The second (boosted) random forest is able to reduce this gap somewhat but not enough. We then look at the proximity score of that anomalous test-point vs the training signal (pseudo-residuals) for the random forests in Figure \ref{fig:poissonproximity}. We've also compared the same for another random point in the test-set to distinguish the effect that the anomalous point has.	
	\afterpage{\clearpage}	
	We see that the outliers have a higher proximity score for the anomaly compared to the random test-point for the first (base) random forest. But for the next step the corresponding training signal (pseudo-residuals) are low - so even though they have similar proximity scores as before they don't influence the prediction too much. For a smaller subsample fraction (0.2) we can see that the proximity scores for all points are much lower and so the outliers don't exert undue influence. As a side-effect the corresponding residuals for the next (boost) step are higher (compared to a higher subsample fraction) but they are not too due high so as to have a bad influence on the MSE after the second step.

	\renewcommand{\thefigure}{\thesubsection.\arabic{figure}}
	
	\section{Further details of empirical studies} \label{sec:moreresults}
	
	In this section we provide further details of our simulation studies from \cref{sec:simresults}. We also discuss how the simulations could be affected if we had a different signal in the link space.
	
	\subsection{More detailed plots} \label{sec:detailedlinearplots}
	
	\setcounter{figure}{0}
	
	We continue with the same model as discussed in \cref{sec:simresults}.	First we show the all the details missing in Figure \ref{fig:testsetplots}, namely how the likelihood and MSE behave for  different values of $M$. We see that the shapes of the plots stay largely the same but the range of values may change depending on $M$. Also in the response scale we show the mean improvement of MSE, instead of the median (still in the pseudo-log scale). This then highlights the effect that outliers can have when considering the MSE, as discussed before in \cref{sec:poissonMSE}.
	
	\begin{figure}[ht]
		\centering
		\caption{Improvements in loglikelihood and MSE (link and response spaces) in the pseudo-log scale}
		\includegraphics[width=\columnwidth]{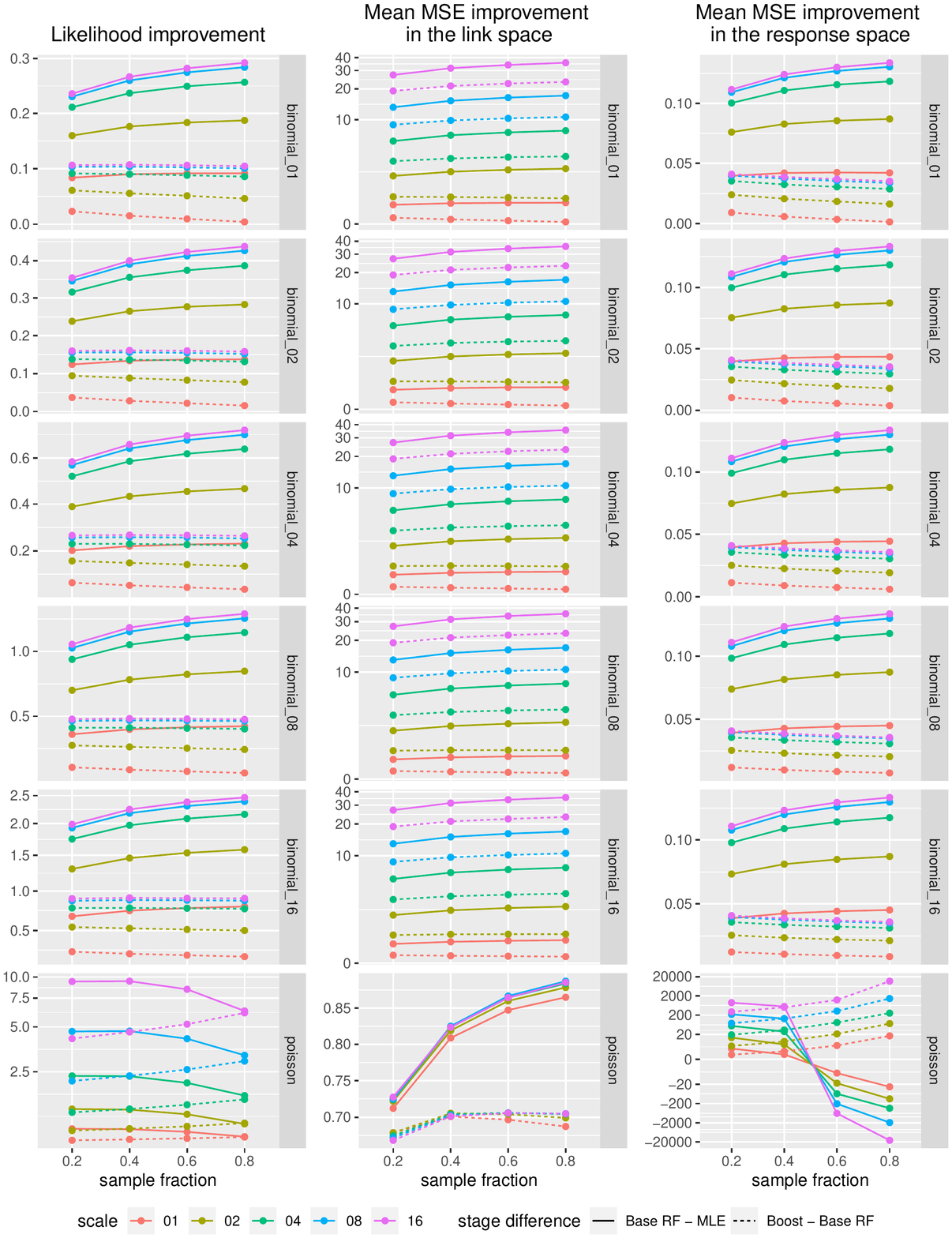}
	\end{figure}
	
	We also include the coverage of the 95\% confidence intervals vs the signals in the response space (that of the link space can be found in Figure \ref{fig:testsetcoverage}). We only provide this for subsample fraction 0.2 but for the others the plots look very similar. We see that the second (boost) random forest improve coverage universally. But the confidence interval has bad coverage for probabilities closer to 0 and 1 - although higher number of trials ($M$) helps the coverage somewhat. In the Poisson case note that the x-axis is in the log-scale. The lower coverage for low values of truth ($\lambda = e^{f(x)}$) may be due to the fact that for low values of $\lambda$ too many $y$'s are zeroes thus making it difficult to distinguish between them, for example $\lambda = 0.1$ vs $\lambda = 0.2$, as well as the truncation effects discussed in Appendix \ref{sec:gbflimits}. 
	
	\begin{sidewaysfigure}[ht]
		\centering
		\caption{Coverage of 95\% confidence intervals vs the true signal in the response space (sample fraction 0.2)}
		\includegraphics[width=\columnwidth]{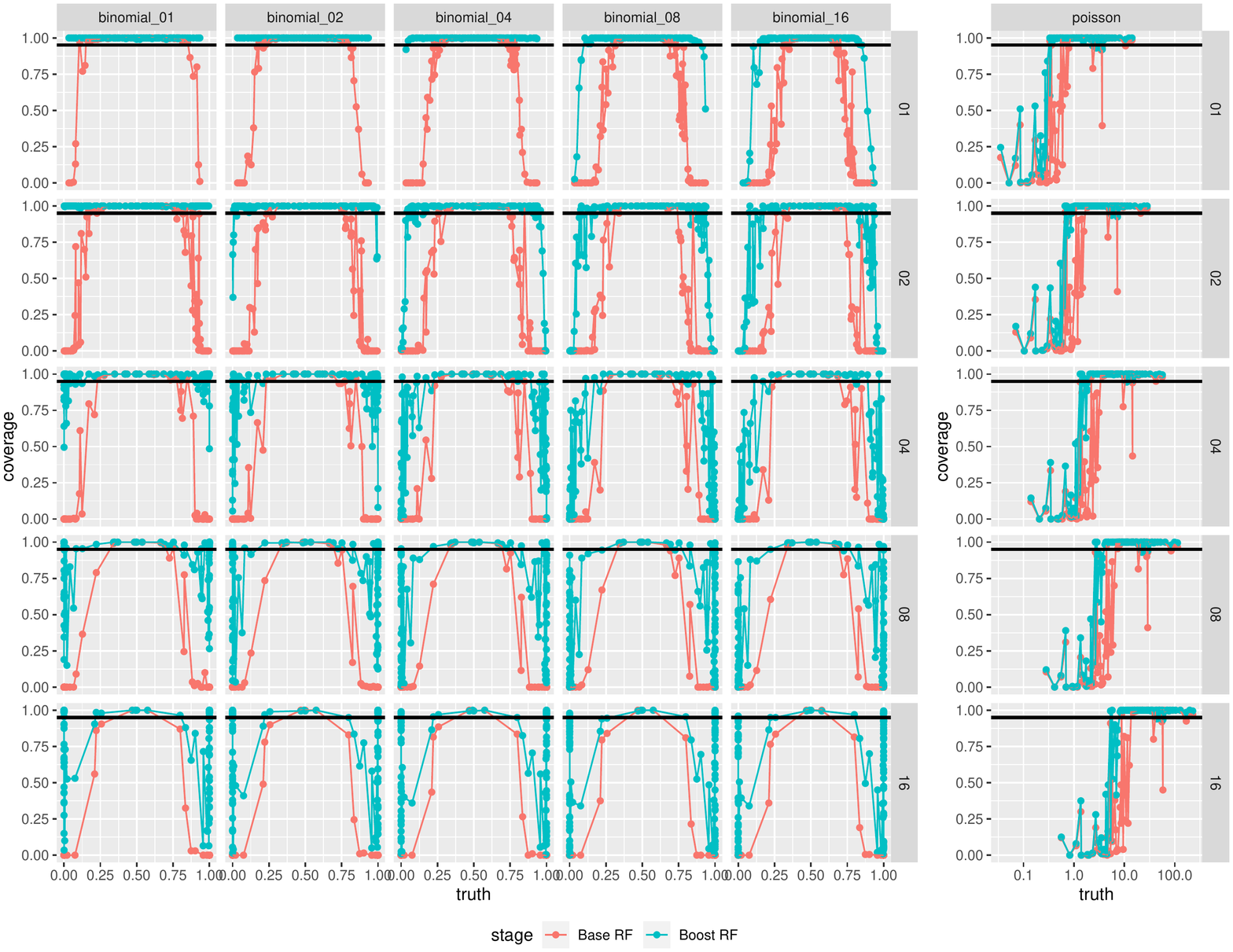}
	\end{sidewaysfigure}
	
	Finally we show more detailed plots and other metrics we considered for the five fixed-points $p_1, \dots, p_5$.
	\begin{itemize}
		\item The absolute bias plots are the same as Figure \ref{fig:5ptsbias} but with more details with all the binomial families (all the value of $M$) and all the 5 points being shown.
		
		\begin{sidewaysfigure}[ht]
			\centering
			\caption{Absolute bias in the link space}
			\includegraphics[width=\textheight]{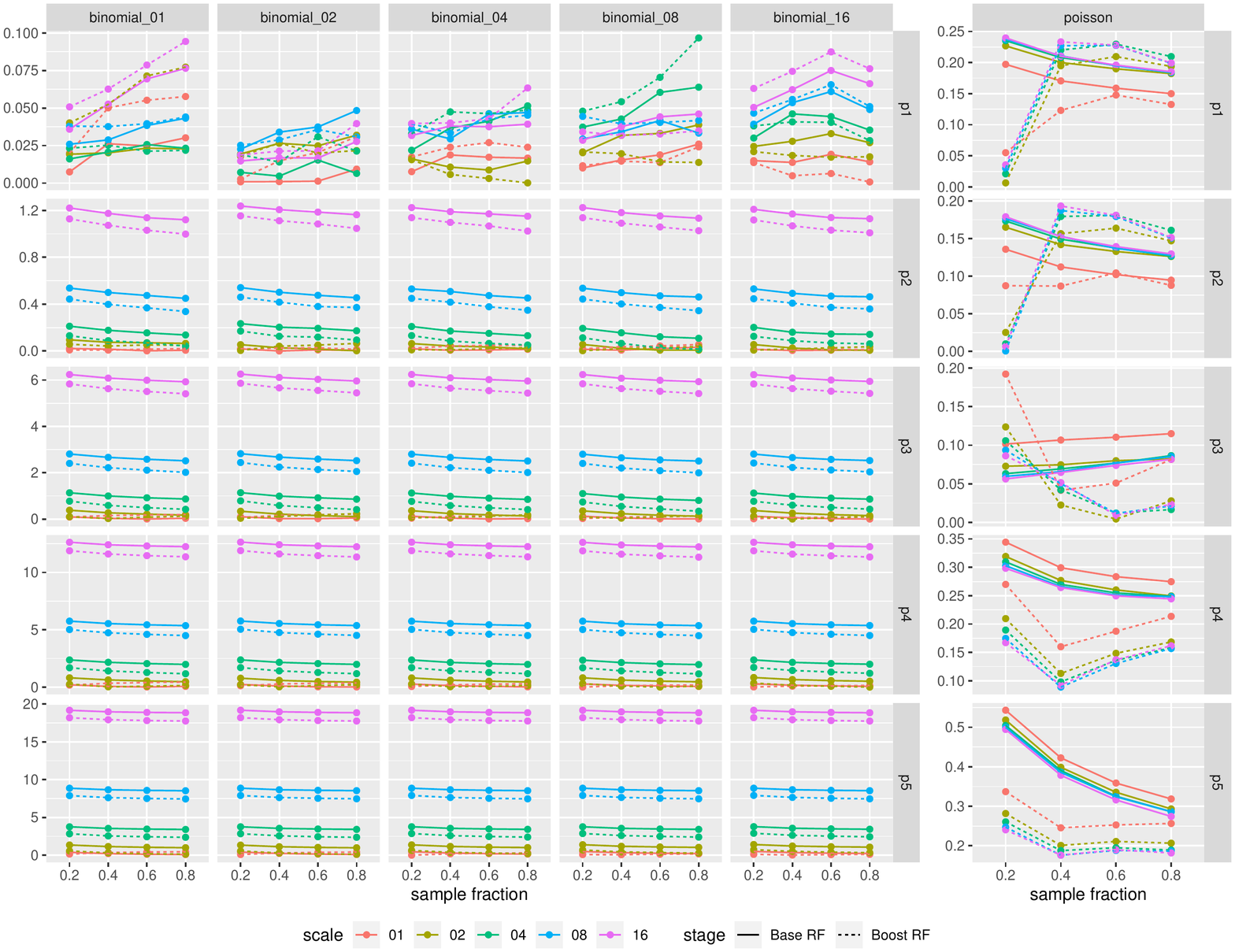}
		\end{sidewaysfigure}
		
		\begin{sidewaysfigure}[ht]
			\centering
			\caption{Absolute bias in the response space}
			\includegraphics[width=\textheight]{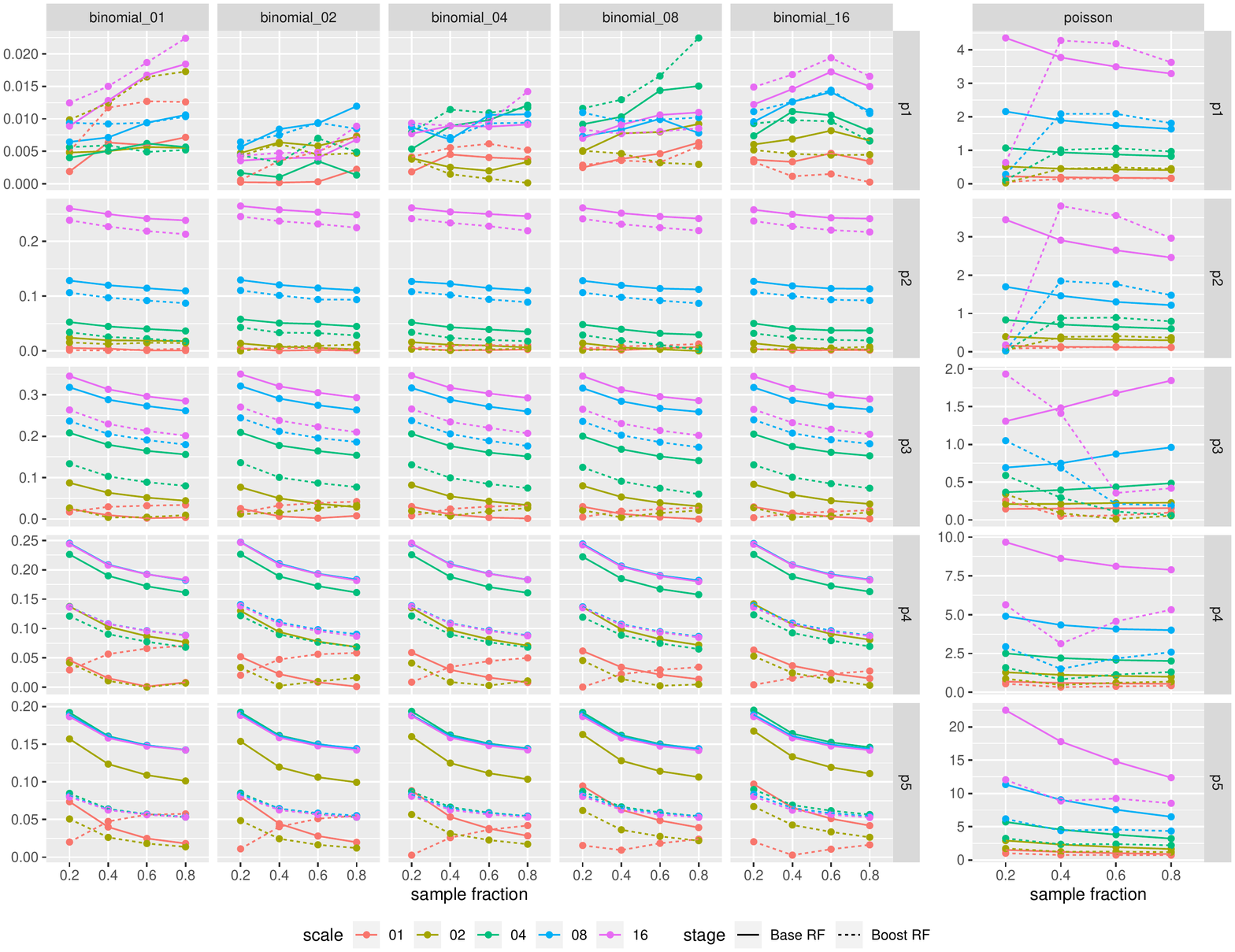}
		\end{sidewaysfigure}
		
		\item The value of the variance estimate is averaged over 200 replicates. For the response space the y-axis is in the log scale. We see that the variance estimate increases after boosting in most of the cases, and it decreases with subsample fraction. Also, at test points further from origin the variance decreases for the binomial cases but increases for the Poisson family. Similar behaviour is also observed for the variance estimate vs scale, increasing scale increases the variance for binomial but decreases for Poisson. Additionally for the binomial cases the variance estimate decreases with the number of trials (value of $M$).
		
		\begin{sidewaysfigure}[ht]
			\centering
			\caption{Average variance estimate in the link space}
			\includegraphics[width=\textheight]{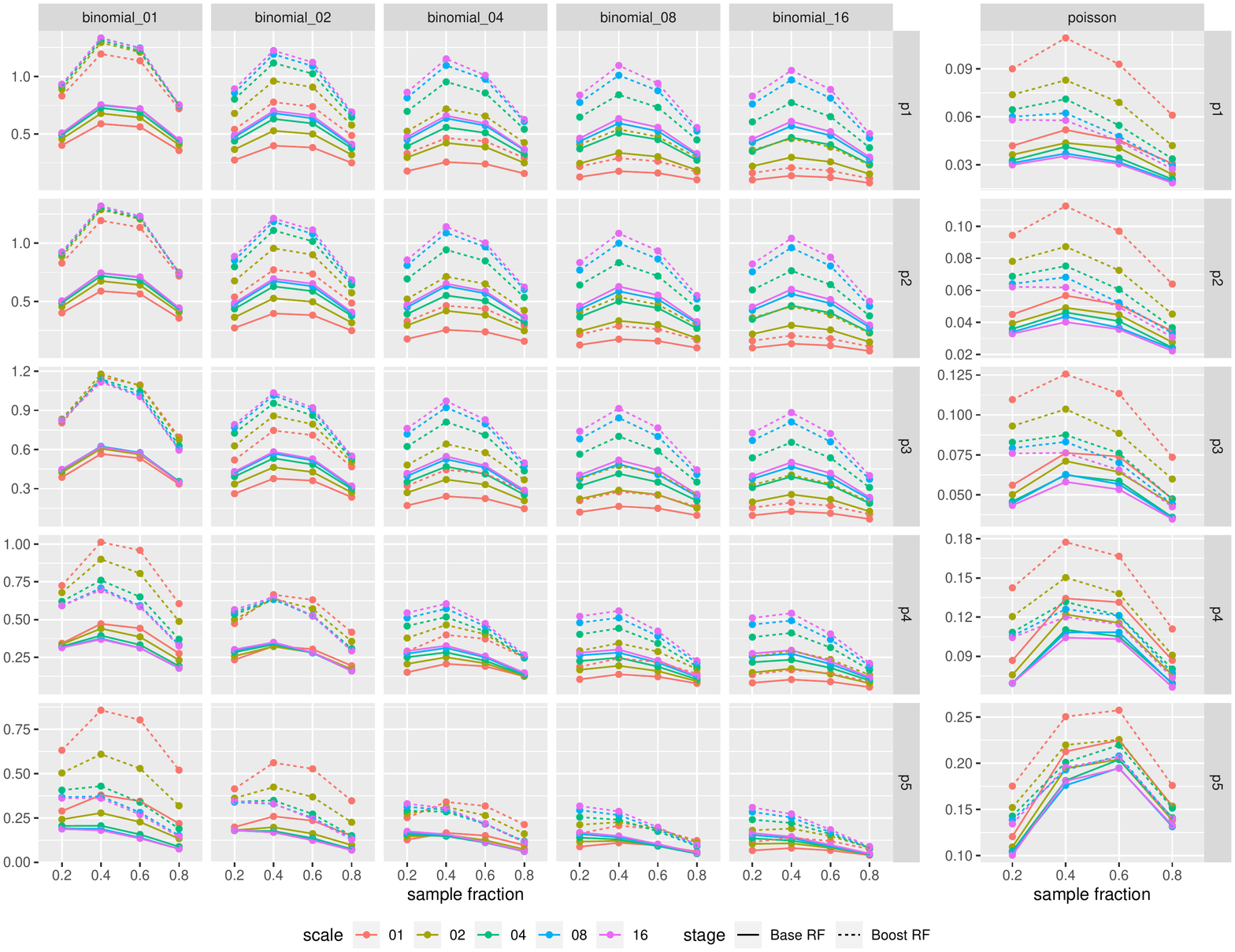}
		\end{sidewaysfigure}
		
		\begin{sidewaysfigure}[ht]
			\centering
			\caption{Average variance estimate in the response space (log scale)}
			\includegraphics[width=\textheight]{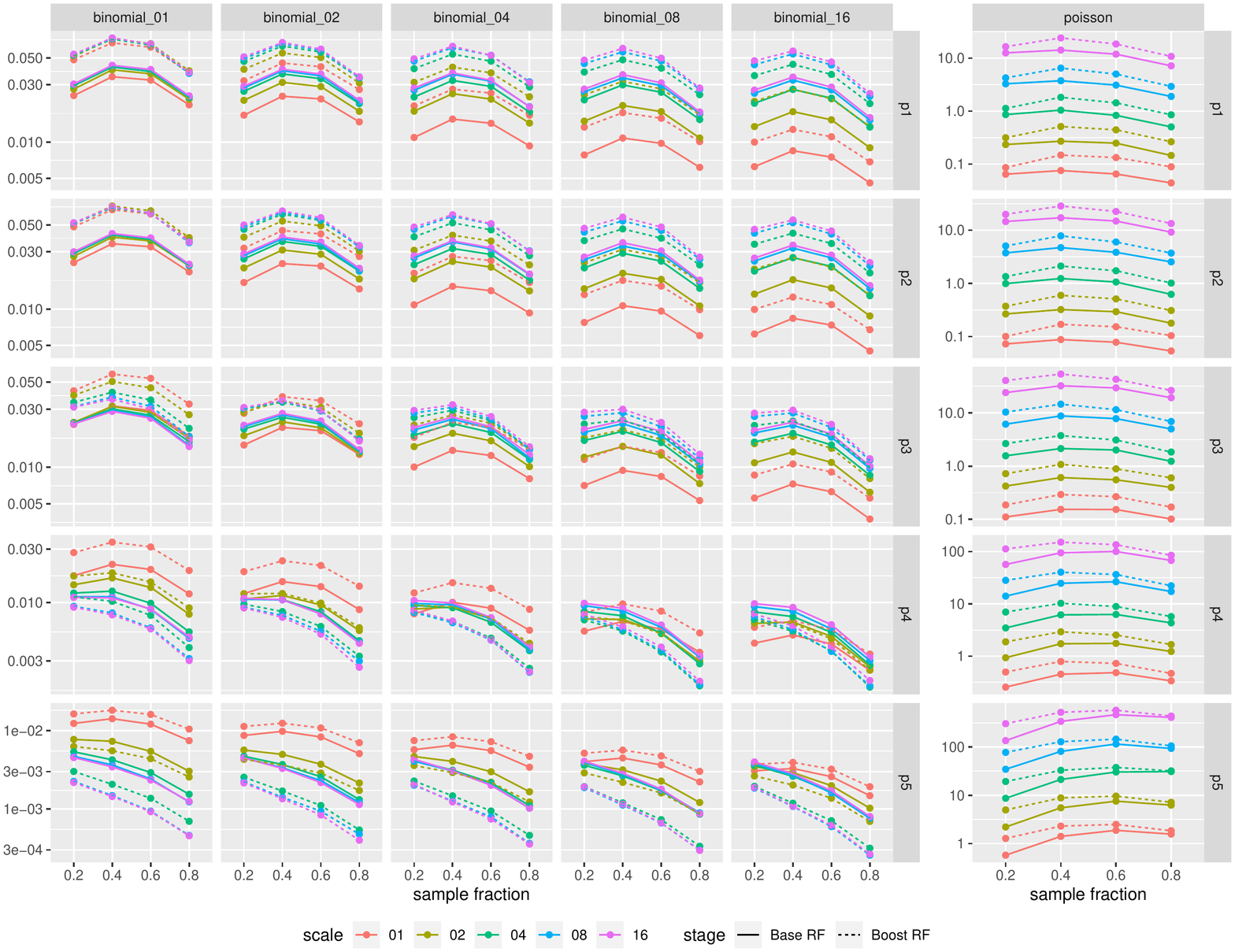}
		\end{sidewaysfigure}
		
		\item The variance consistency Figure for the link space is the same as Figure \ref{fig:5ptsvarcons} but with more details. We don't observe much of a change in the consistency across different value of the $M$. For the response space the consistency is presented in the log-scale. Note that for the binomial cases the actual values of the variance and its estimate are  small so fluctuations can result in high ratios. On the other hand we see that the variance is very consistent for the Poisson case.
		
		\begin{sidewaysfigure}[ht]
			\centering
			\caption{Variance consistency in the link space}
			\includegraphics[width=\textheight]{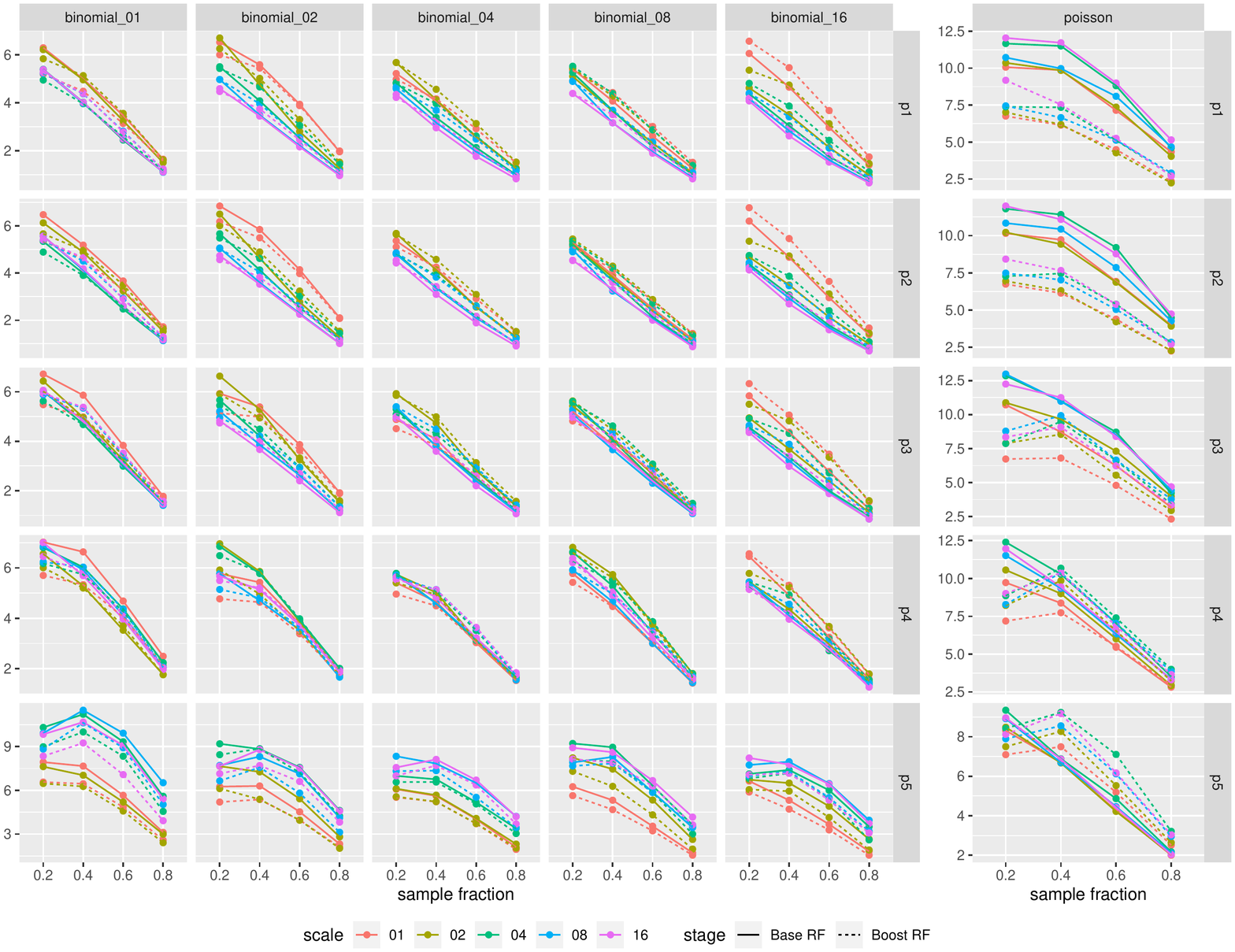}
		\end{sidewaysfigure}
		
		\begin{sidewaysfigure}[ht]
			\centering
			\caption{Variance consistency in the response space (log scale)}
			\includegraphics[width=\textheight]{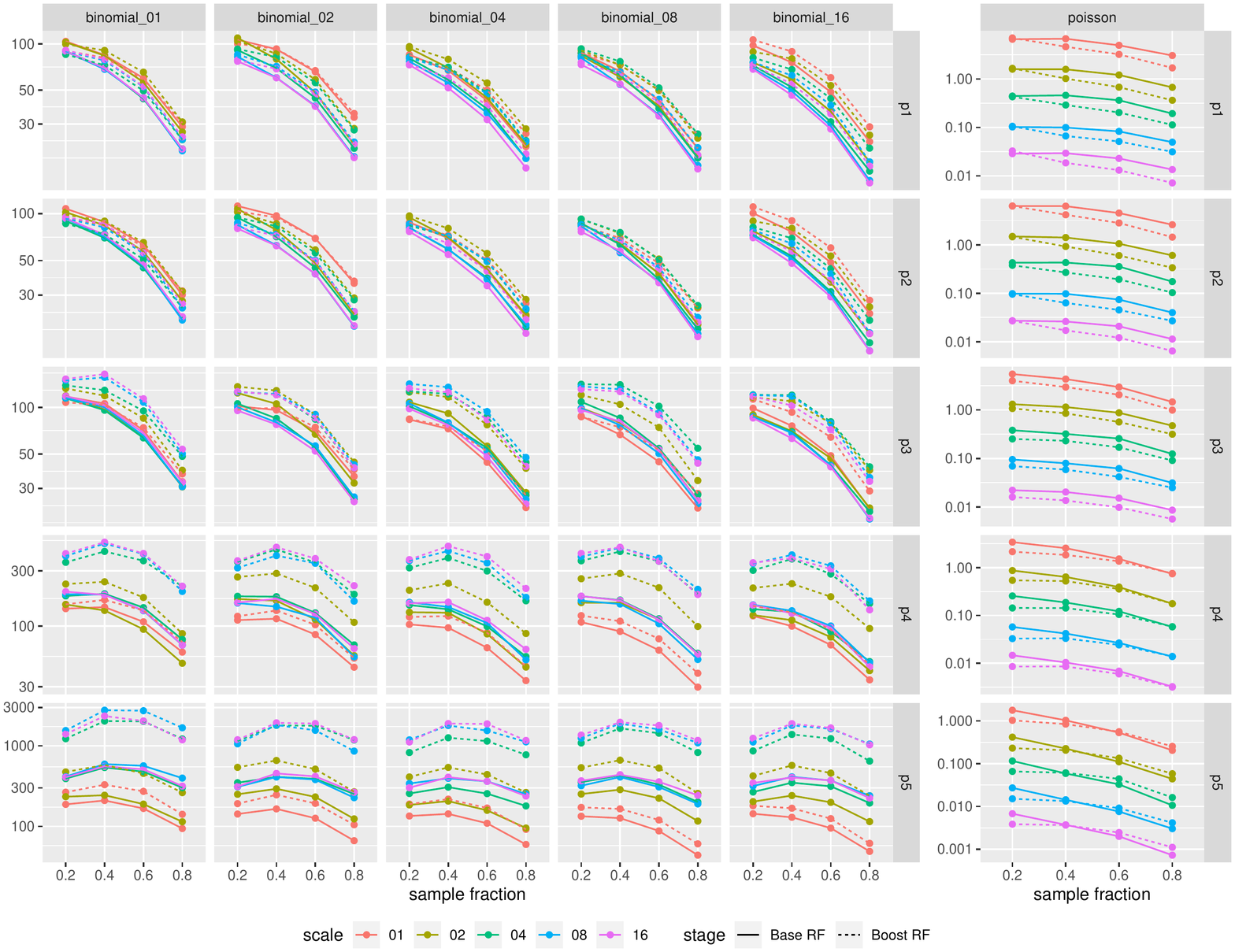}
		\end{sidewaysfigure}
		
		\item The Kolmogorov-Smirnov statistic for normality is always pretty small for the link space and thus we can conclude that the asymptotic normality mentioned in \ref{thm:main} probably holds true. For the response space the statistic is generally larger as might be expected from nonlinear transformations. We also see that the K-S statistic is large for the boosted stage compared to the base random forest and also for larger values of subsample fraction. It's relationship with number of trials (in the binomial case) and scale (in both families) seems mixed.
		
		\begin{sidewaysfigure}[ht]
			\centering
			\caption{K-S statistic for normality in the link space}
			\includegraphics[width=\textheight]{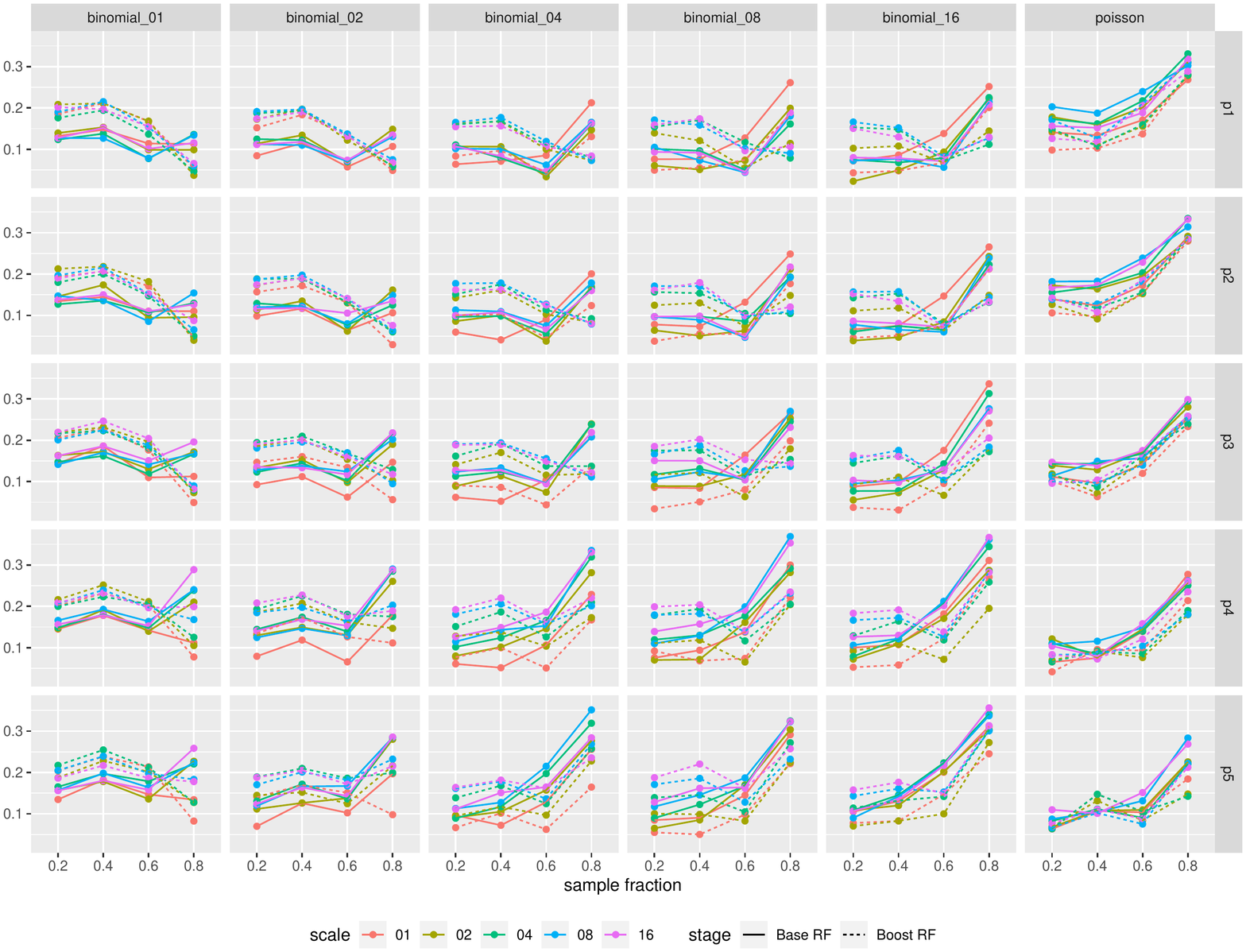}
		\end{sidewaysfigure}
		
		\begin{sidewaysfigure}[ht]
			\centering
			\caption{K-S statistic for normality in the response space}
			\includegraphics[width=\textheight]{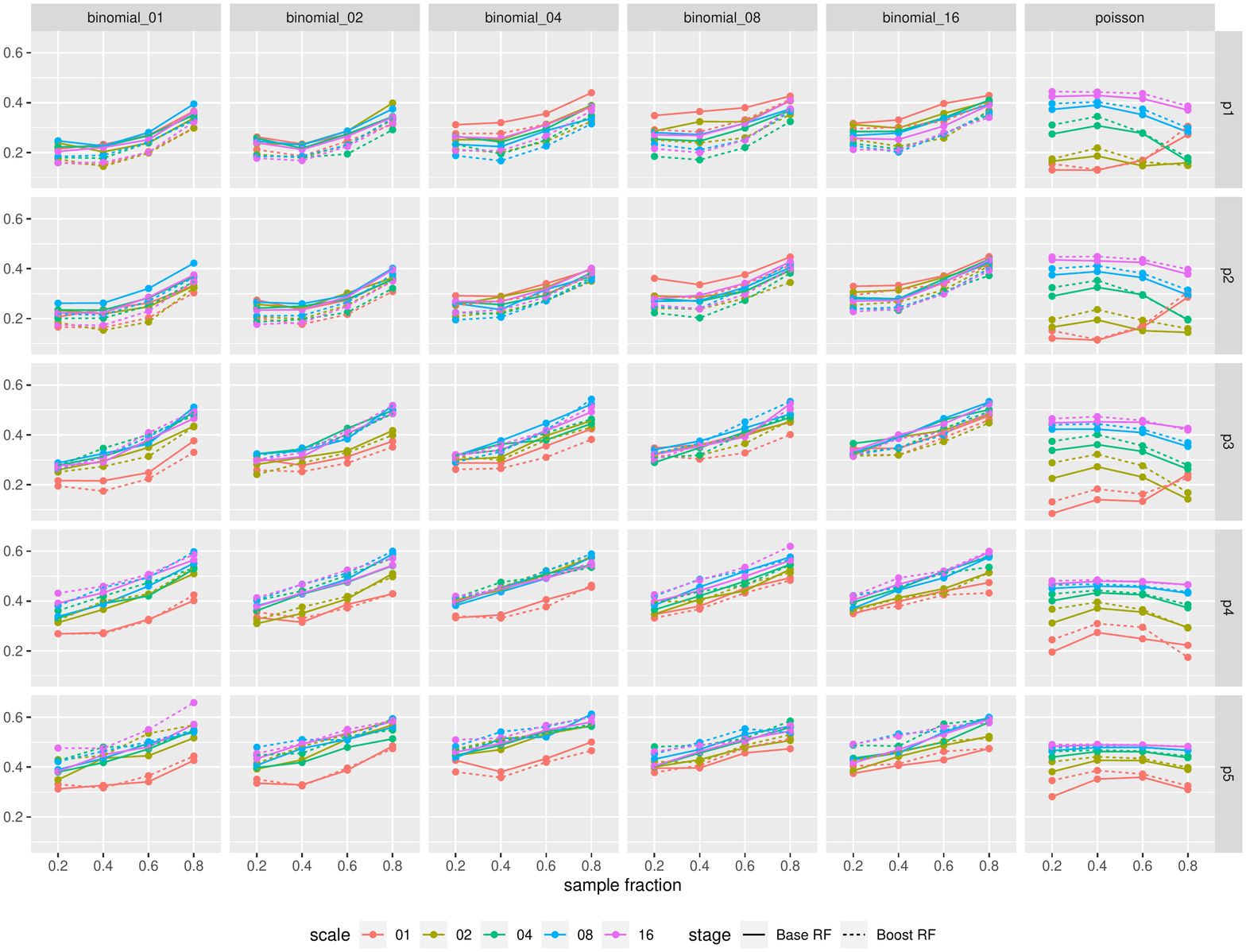}
		\end{sidewaysfigure}

		\item The MSE is averaged over the 200 replicates and shown in the log-scale. We see that the MSE for points nearer the origin is higher after the boost stage than after the base stage, although this behaviour is reversed for points farther away from the origin. The MSE also increases for farther away points, and also mostly as the subsample fraction increases. Higher scales also have higher MSE in general. There doesn't seem to be any relationship of MSE with number of trials in the binomial case.
		
		\begin{sidewaysfigure}[ht]
			\centering
			\caption{Average MSE in the link space (log scale)}
			\includegraphics[width=\textheight]{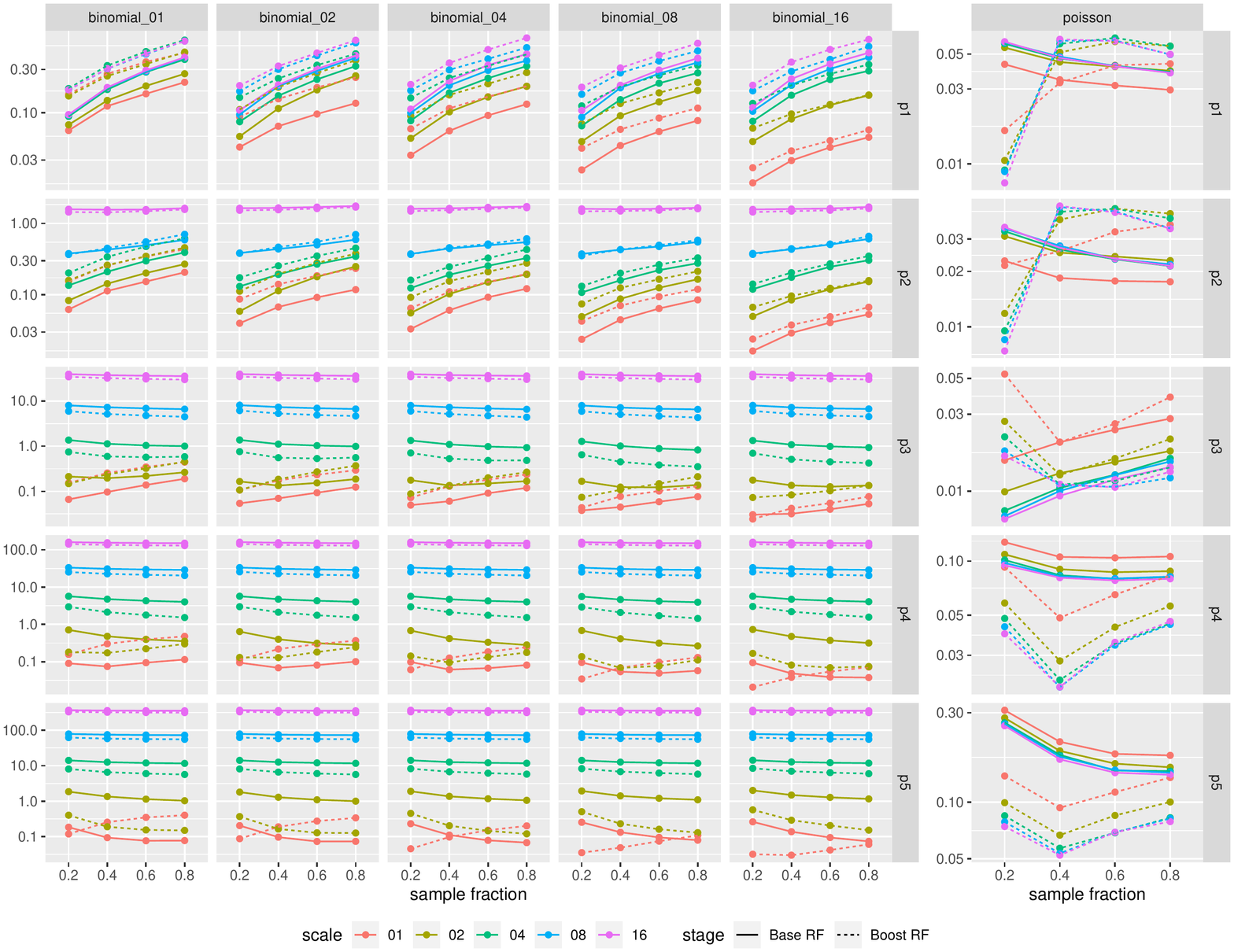}
		\end{sidewaysfigure}
		
		\begin{sidewaysfigure}[ht]
			\centering
			\caption{Average MSE in the response space (log scale)}
			\includegraphics[width=\textheight]{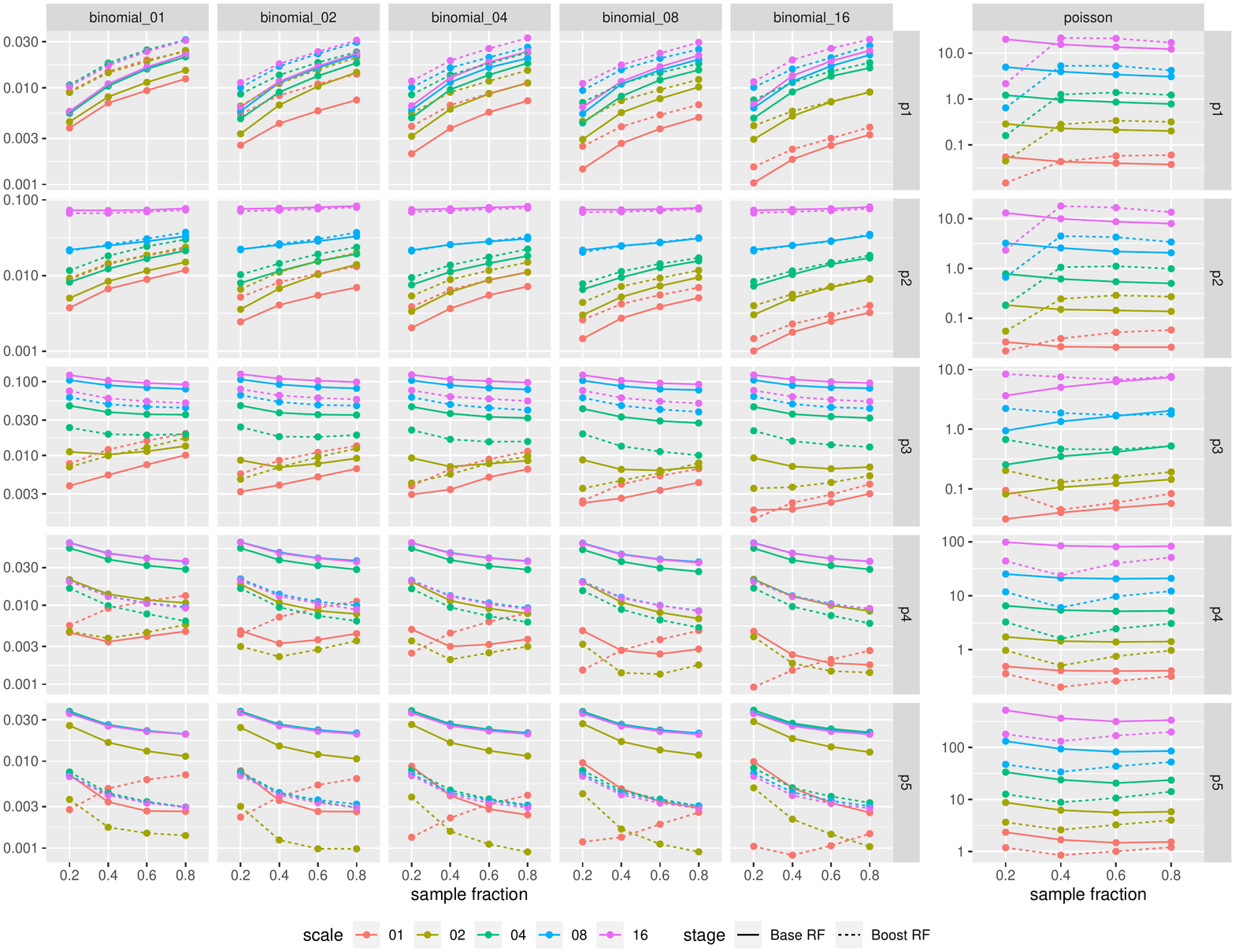}
		\end{sidewaysfigure}
		
		\afterpage{\clearpage}
	\end{itemize}

	\subsection{Non-linear signal} \label{sec:nonlinearplots}
	
	\setcounter{figure}{0}
	
	The model discussed here is exactly the same as in \cref{sec:simresults} except that the signal is $f(x) = \|x\|_2 - \frac{\sqrt{m}}{2}$ instead of $f(x) = \sum_{i=1}^5 x_i$. $\frac{\sqrt{m}}{2}$ is subtracted to have some of the link space signals be negative. 
		
	The likelihood and MSE always improves for both stages. This improvement is generally better with scale and slightly better with subsample fraction. The number of trials in the binomial case doesn't seem to have an effect. Most notably the effect of outliers as discussed in \cref{sec:simresults} seem to be non-existent in the Poisson case. This is probably because for test-points far away from the origin the signal (norm) smooths out - so the ``outliers'' defined in the sense of far-away points don't actually have extreme signals with which to affect the residuals and/or predictions. Thus adding random forests always seems to improve the MSE in the response space.
	
	\begin{figure}[ht]
		\centering
		\caption{Improvements in loglikelihood and MSE (link and response spaces) in the pseudo-log scale}
		\includegraphics[width=\columnwidth]{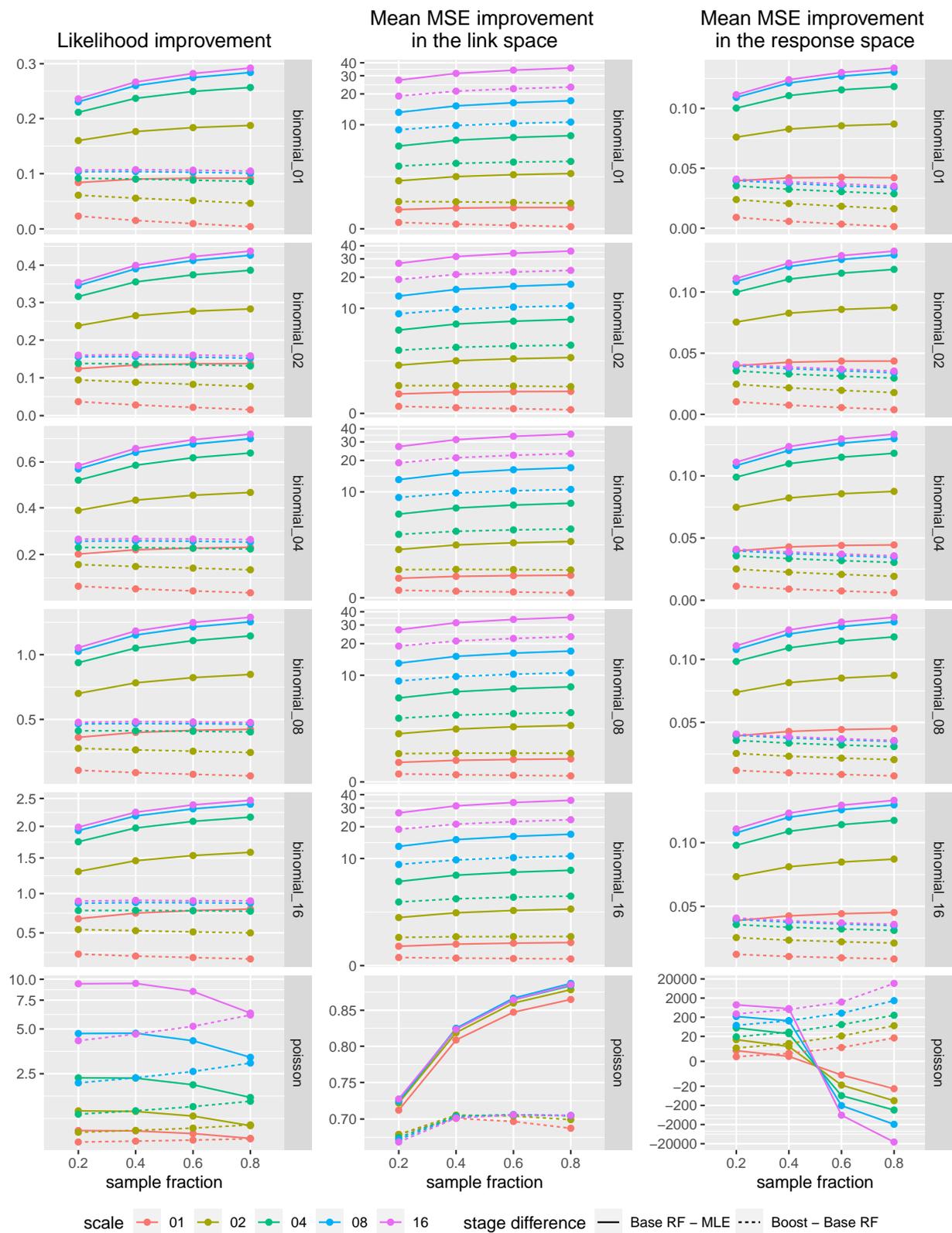}
	\end{figure}
	
	Next we look at coverage of the 95\% confidence interval vs the true value of the signal (pseudo-log scale). We present plots for only one subsample fraction as the other ones look very similar. In the link space, similar to the case for the linear signal (figure \ref{fig:testsetcoverage}), the boost step improves coverage everywhere and the extreme values have zero coverage. Also the range of true signal with good coverage is higher compared to Figure \ref{fig:testsetcoverage}. Note that the norm function increases rapidly away from the origin at first, i.e, has a high gradient but then plateaus farther away. This is why lower values of the true signal are harder to estimate consistently than higher values. Thus negative values of $f(x)$ in the link space have lower confidence coverage than the corresponding positive values (for example -2 vs 2); the same thing can be observed for probabilities lower than 0.5 for the binomial cases in the response space. For the Poisson family in the response space higher values of the true signal have low coverage for a related reason - since higher value of the norm have low gradient, then the particular true signal gets harder to pin down in the confidence interval, something which becomes worse with higher scales. Finally the number of trials in the binomial cases (values of $M$) seems to refine the confidence coverage somewhat although this effect is diminished with scale.
	
	\begin{sidewaysfigure}[ht]
		\centering
		\caption{Coverage of 95\% confidence intervals vs the true signal in the link space (sample fraction 0.4)}
		\includegraphics[width=\columnwidth]{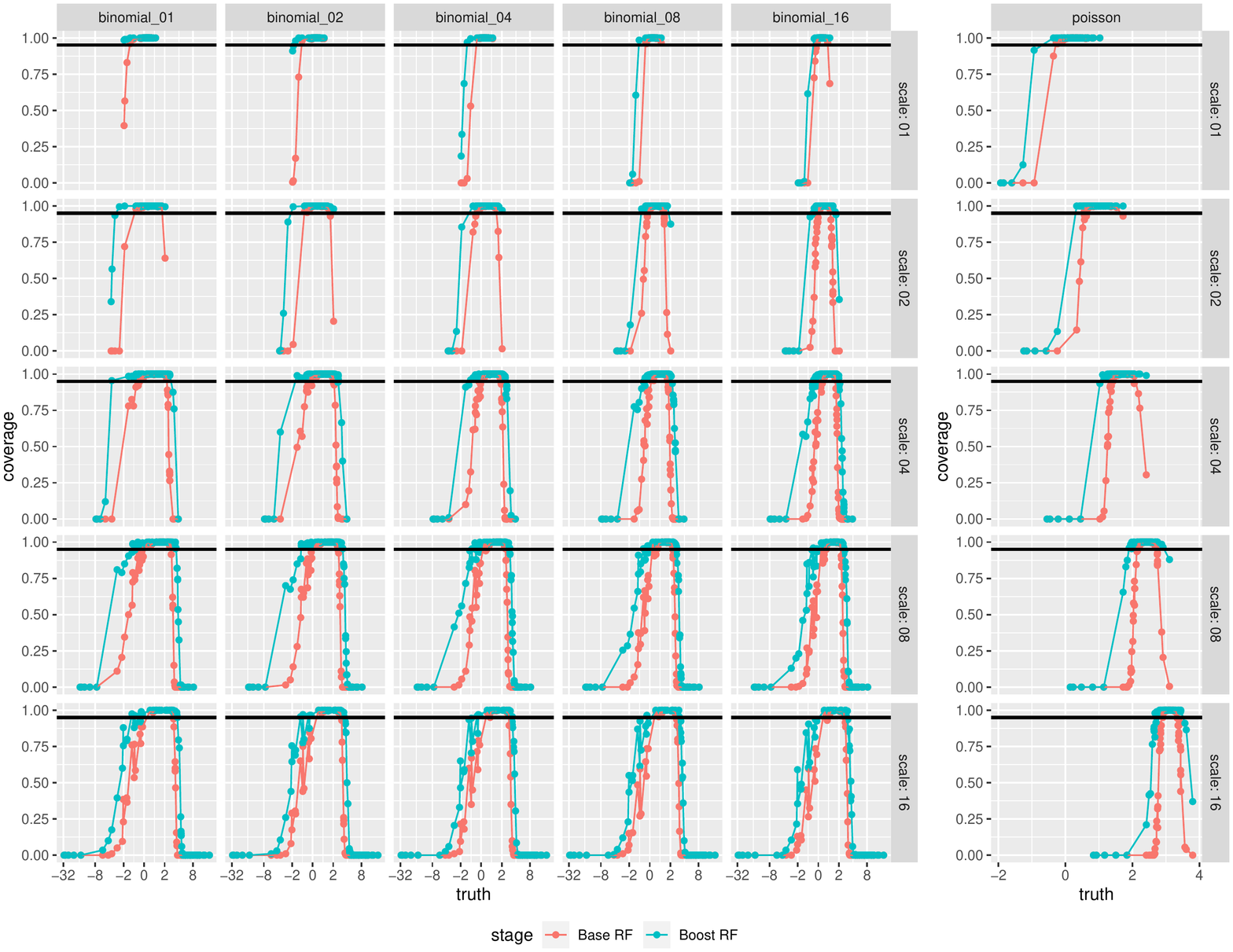}
	\end{sidewaysfigure}
	
	\begin{sidewaysfigure}[ht]
		\centering
		\caption{Coverage of 95\% confidence intervals vs the true signal in the response space (sample fraction 0.2)}
		\includegraphics[width=\columnwidth]{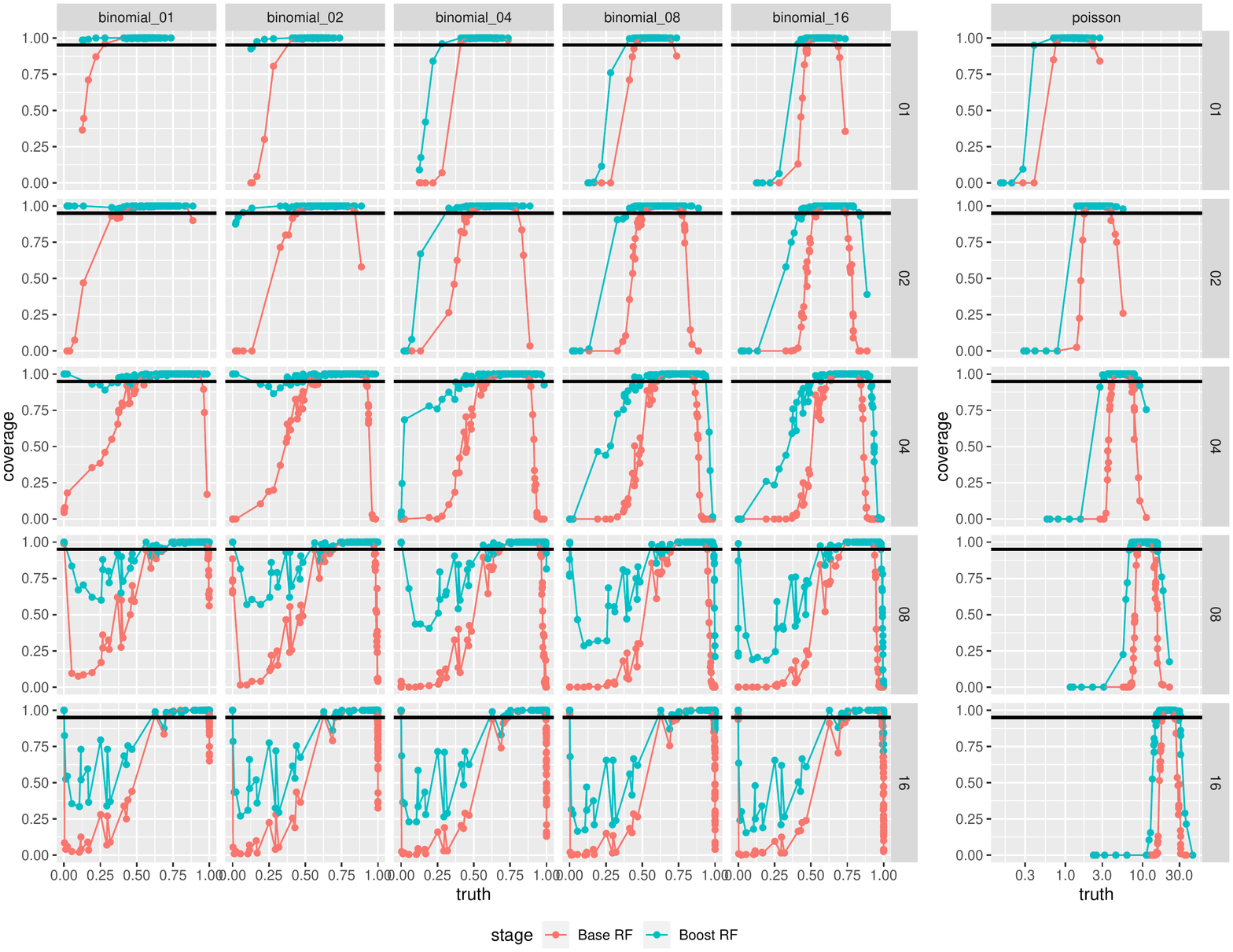}
	\end{sidewaysfigure}
	
	We then show plots corresponding to the 5 fixed points for all the metrics as discussed in \cref{sec:detailedlinearplots}
	
	\begin{itemize}
		\item Absolute bias decreases for the boosted stage as compared to just the base random forest. It also decreases as the points get farther from the origin - although this effect is much lower in the response space. Bias also decreases as subsample fraction increases - most pronounced for the binomial cases in the response space. The effect of scale is slightly inconclusive - for the binomial cases in the response space the absolute bias increases with scale first before subsidising - for all other cases bias increases with scale. The number of trials in the binomial cases do not seem to have an effect.
		
		\begin{sidewaysfigure}[ht]
			\centering
			\caption{Absolute bias in the link space (log scale)}
			\includegraphics[width=\textheight]{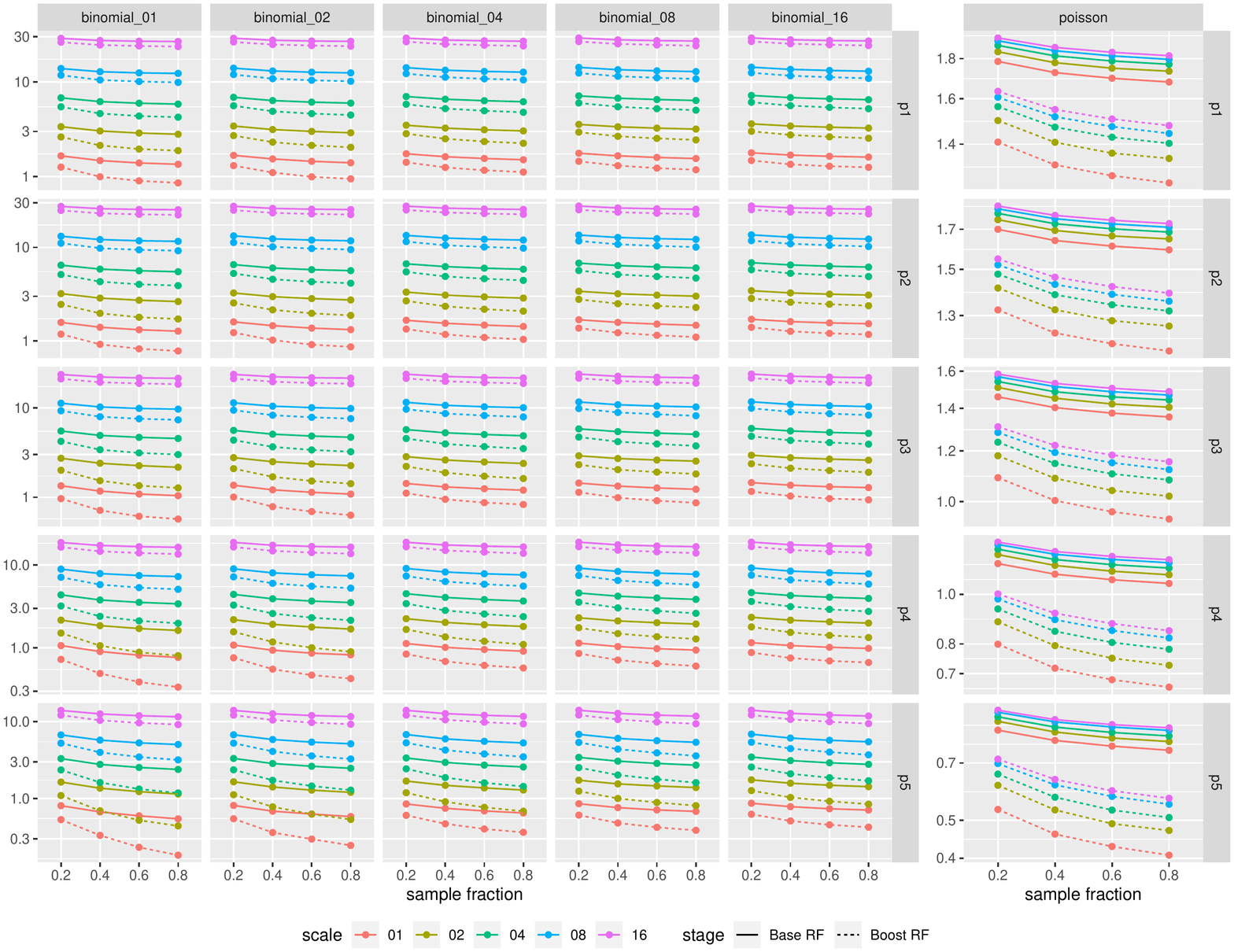}
		\end{sidewaysfigure}
		
		\begin{sidewaysfigure}[ht]
			\centering
			\caption{Absolute bias in the response space (Poisson family in log scale)}
			\includegraphics[width=\textheight]{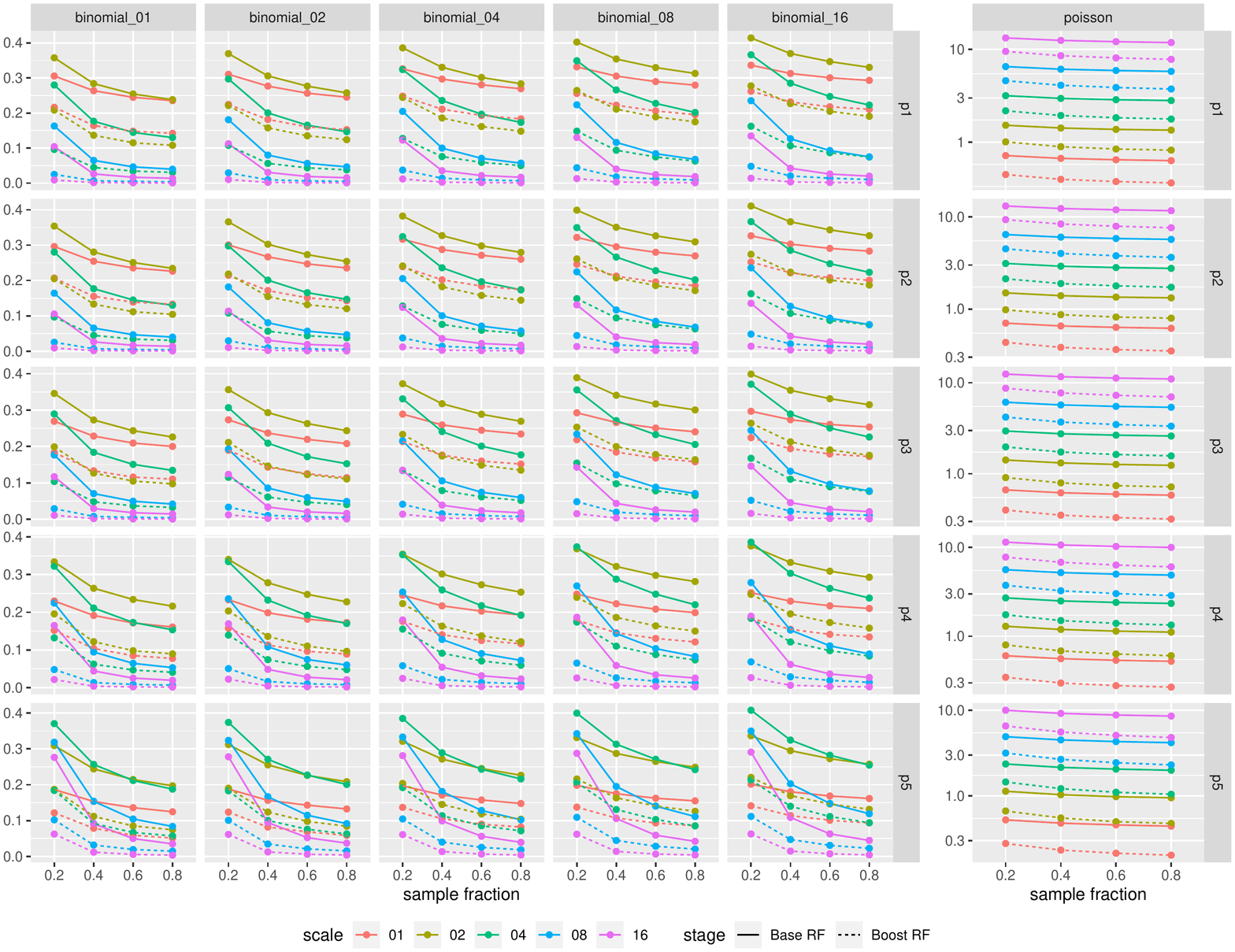}
		\end{sidewaysfigure}
		
		\item The average variance estimate is higher after the second (boosted) stage than the first stage and it also seems to increase with subsample size first and then decreases - except for the binomial cases in the in the response space where the boosted stage can sometimes have lower variance than the base stage and the variance may also consistently decrease with subsample fraction. The effect of scale is also mixed - for the binomial cases in the link space scale increases variance but in the response space it is the opposite; for the Poisson family we see the reverse, i.e, scale increases variance in the response space but it decreases with scale in the link space. Variance seems to increase slightly as the points get farther away from the origin and the number of trials (for binomials) slightly decreases variance in the link space but doesn't seem to have any effect in the response space.
		
		\begin{sidewaysfigure}[ht]
			\centering
			\caption{Average variance estimate in the link space (log scale)}
			\includegraphics[width=\textheight]{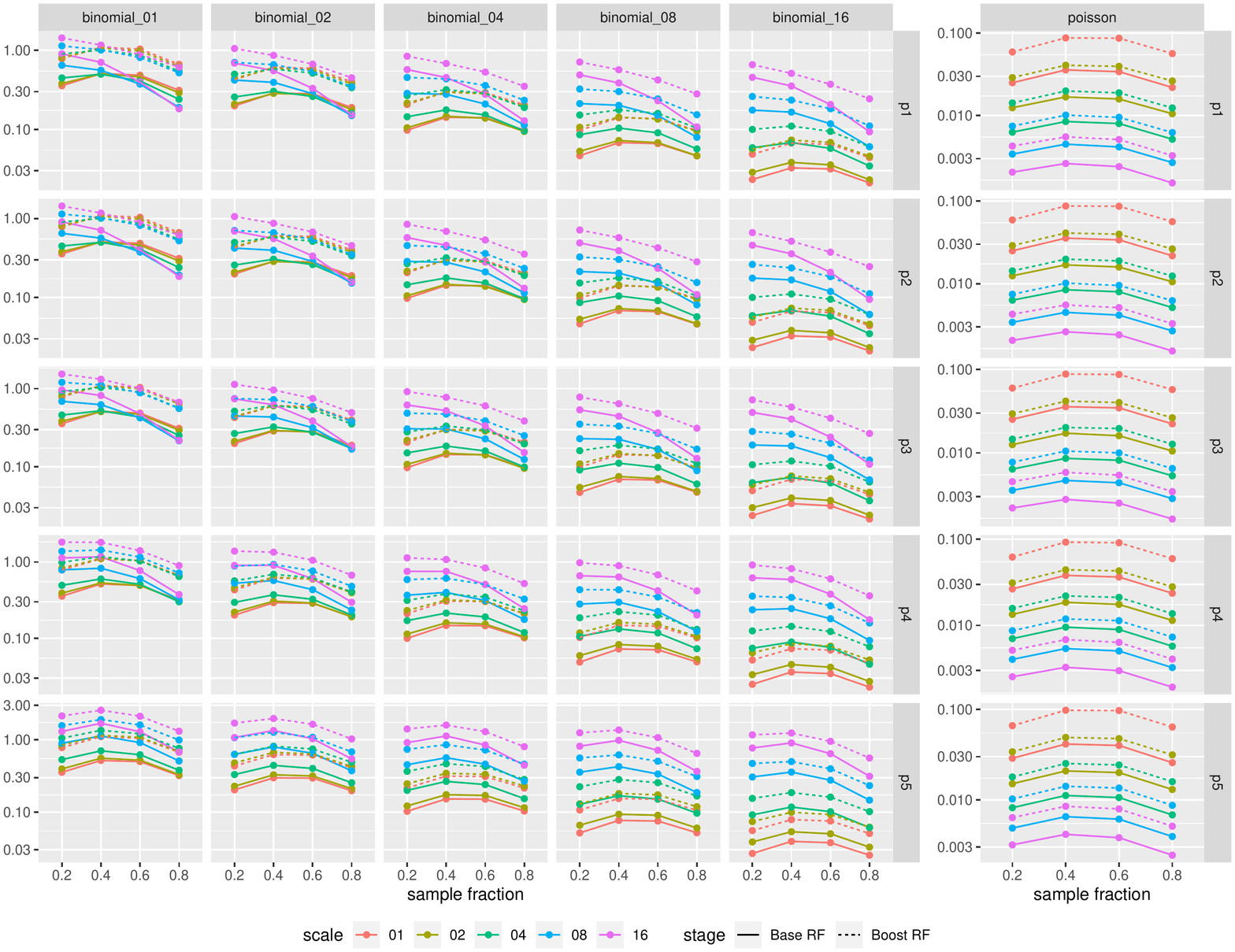}
		\end{sidewaysfigure}
		
		\begin{sidewaysfigure}[ht]
			\centering
			\caption{Average variance estimate in the response space (log scale)}
			\includegraphics[width=\textheight]{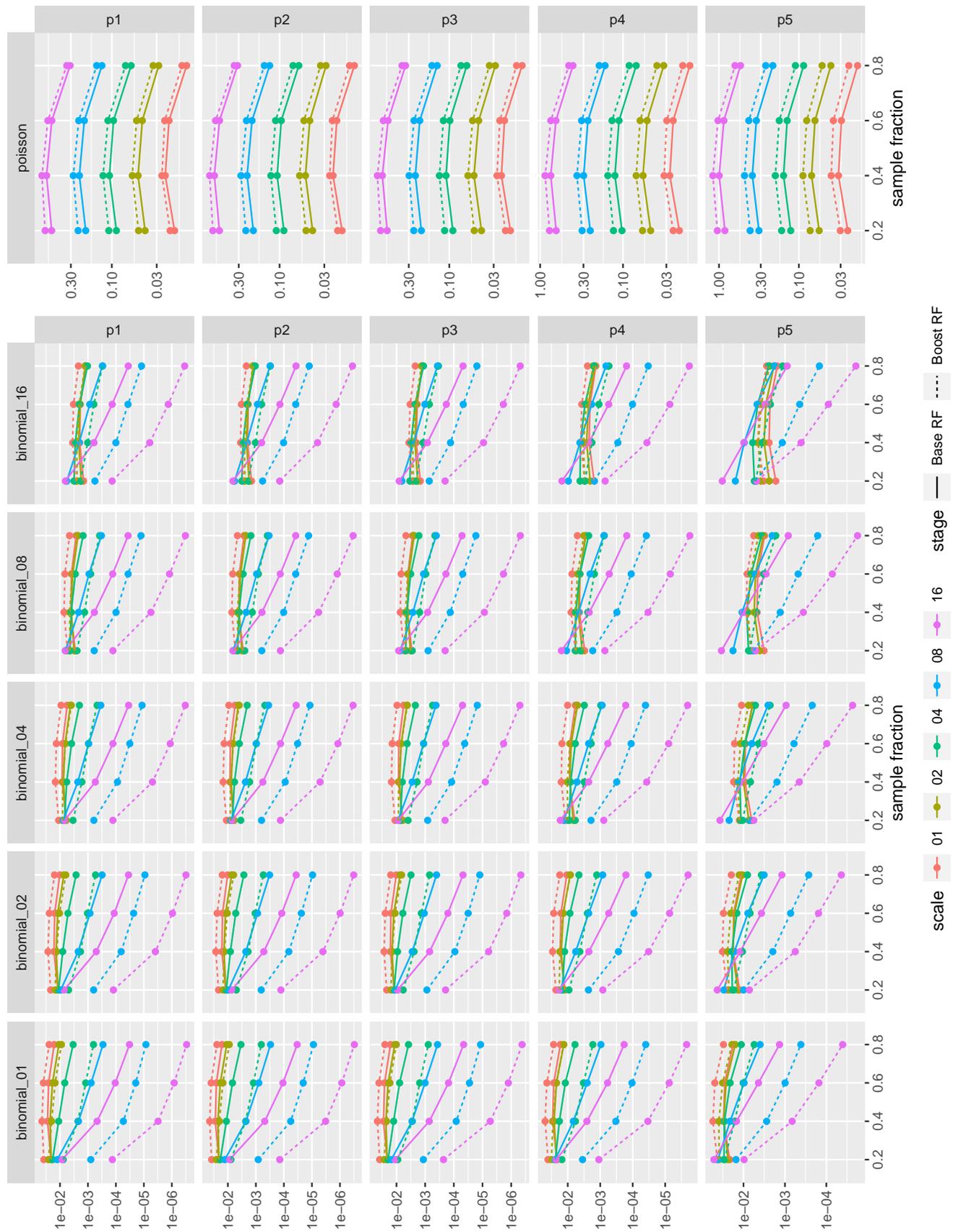}
		\end{sidewaysfigure}
		
		\item The plots for variance consistency shows that our variance estimate is more consistent as scale and subsample fraction increases - except for the binomial case in the response space where both effects seems to be the opposite. But as discussed before in \cref{sec:detailedlinearplots} the quantities we're comparing for the variance consistency are both very small so the results could be unstable. The variance estimate also seems to be more consistent for the boosted stage in the link space and for the base stage in the response space. The consistency metric doesn't seem to change too much as the testpoints move away from the origin. Finally increasing the number of trials in the binomial cases makes the variance estimate a little bit more consistent in the link space but doesn't seem to have any effect in the response space.
		
		\begin{sidewaysfigure}[ht]
			\centering
			\caption{Variance consistency in the link space}
			\includegraphics[width=\textheight]{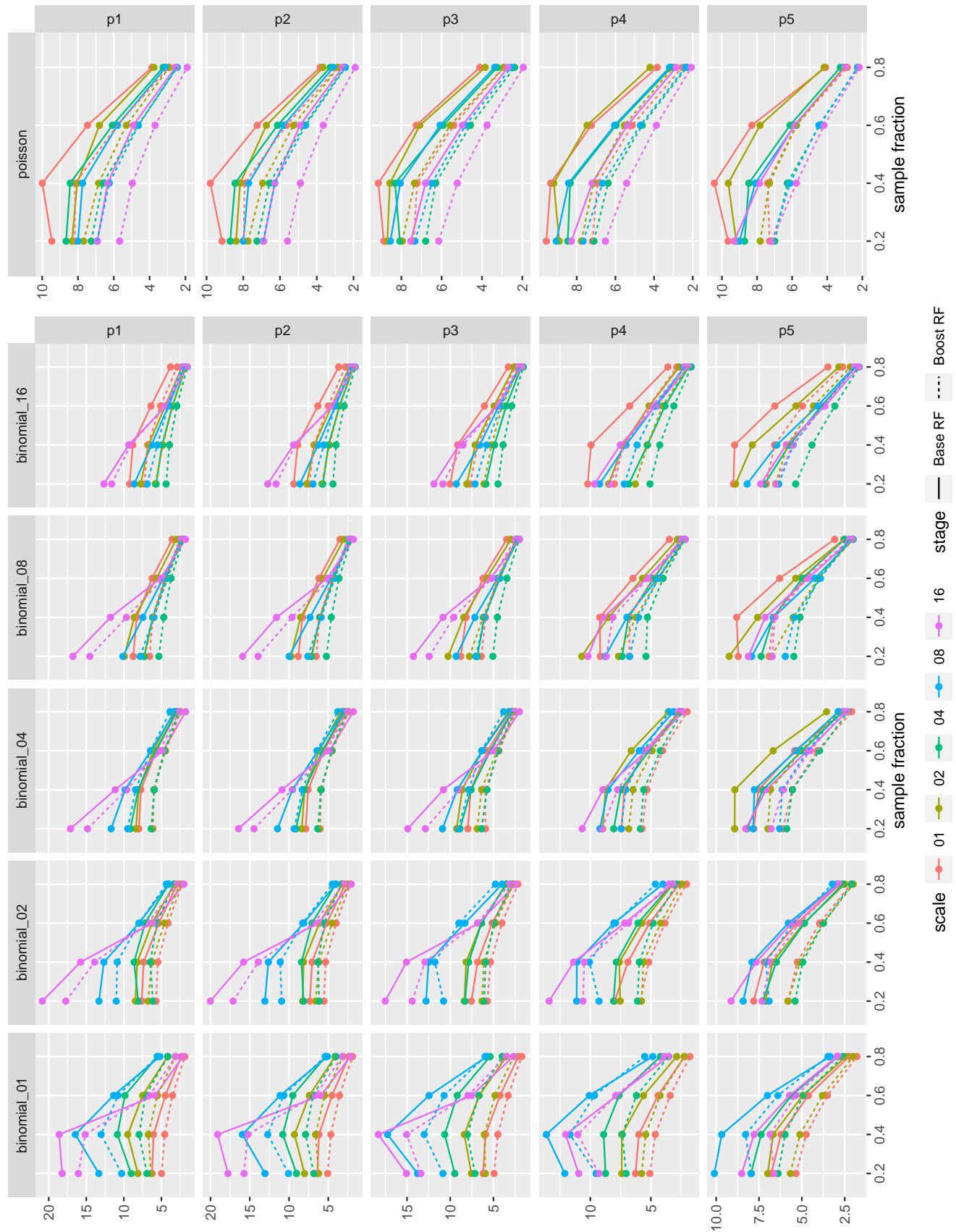}
		\end{sidewaysfigure}
		
		\begin{sidewaysfigure}[ht]
			\centering
			\caption{Variance consistency in the response space (log scale)}
			\includegraphics[width=\textheight]{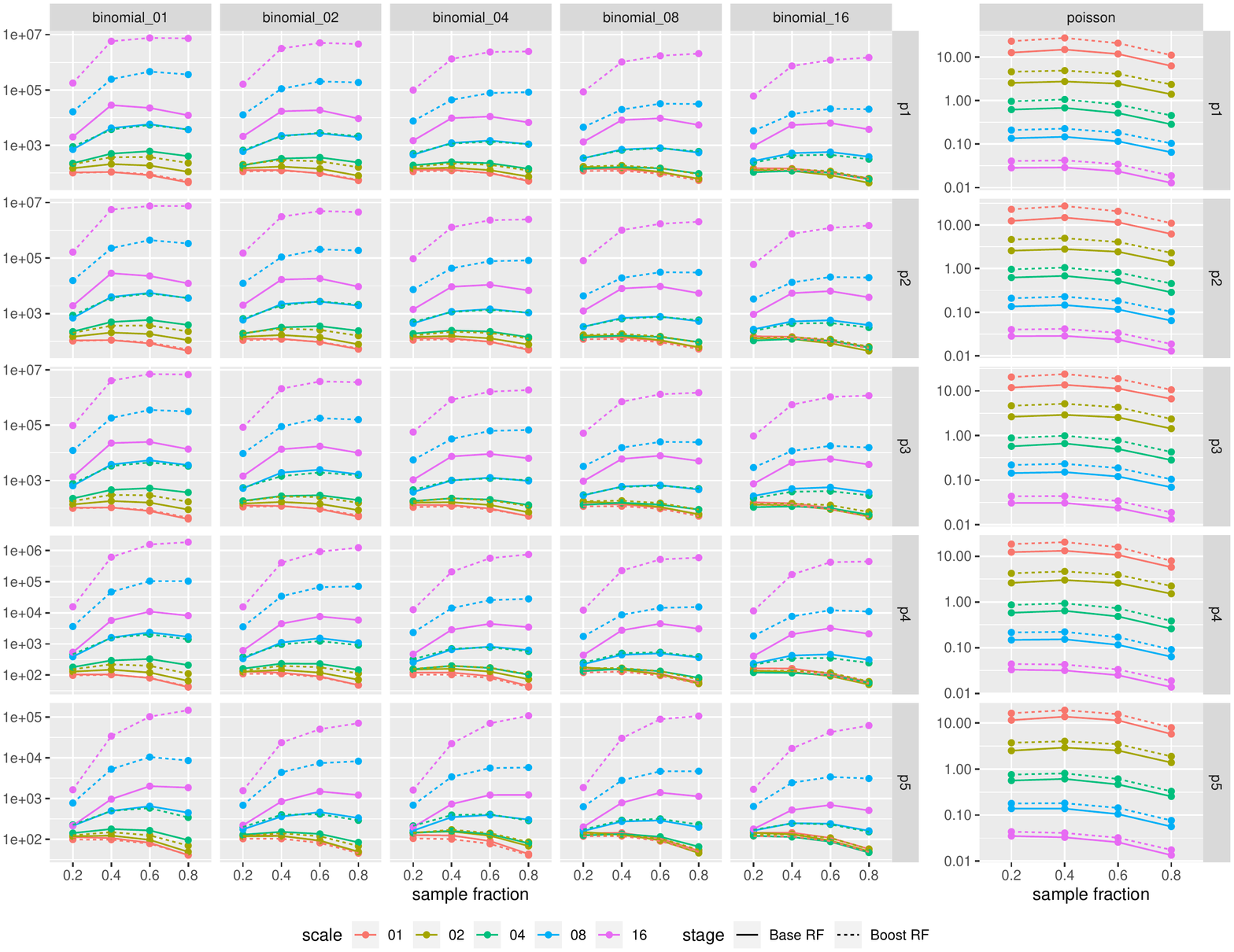}
		\end{sidewaysfigure}
		
		\item The Kolmogorov-Smirnov statistic for normality has reasonable values - it seems to increase slightly with subsample fraction and scale. It has a higher value after the boosted stage except for the Poisson family in the link space where the base stage has a higher K-S statistic. The distance of the test point from the origin doesn't seem to have an effect, and neither does the number of trials in the binomial case.
		
		These statistics are always pretty small for the link space and thus we can conclude that asymptotic normality mentioned in \ref{thm:main} is reasonable. For the response space the statistic is generally larger. We also see that the K-S statistic is large for the boosted stage compared to the base random forest and also for larger values of subsample fraction. It's relationship with number of trials (in the binomial case) and scale (in both families) seems mixed.
		
		\begin{sidewaysfigure}[ht]
			\centering
			\caption{K-S statistic for normality in the link space}
			\includegraphics[width=\textheight]{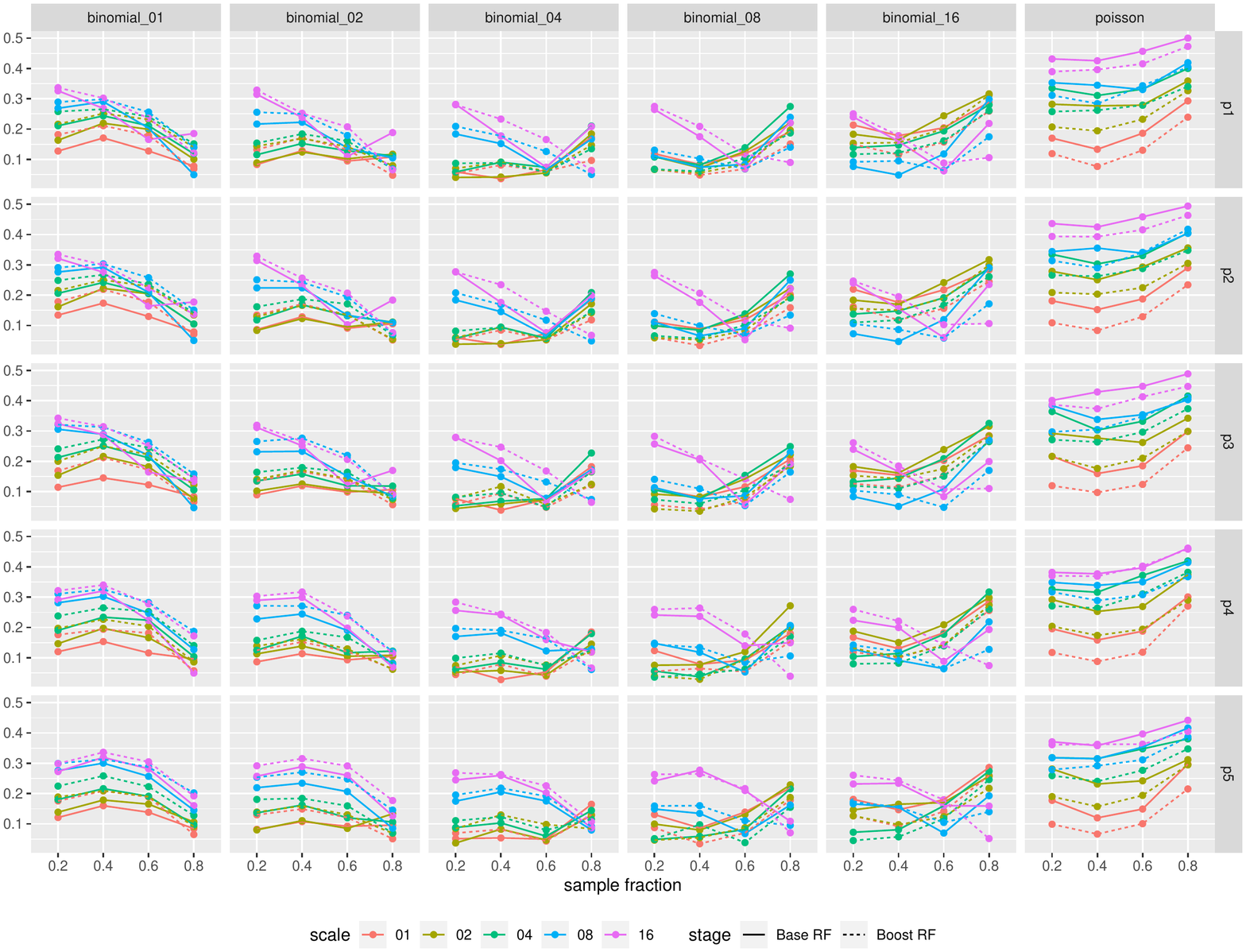}
		\end{sidewaysfigure}
		
		\begin{sidewaysfigure}[ht]
			\centering
			\caption{K-S statistic for normality in the response space}
			\includegraphics[width=\textheight]{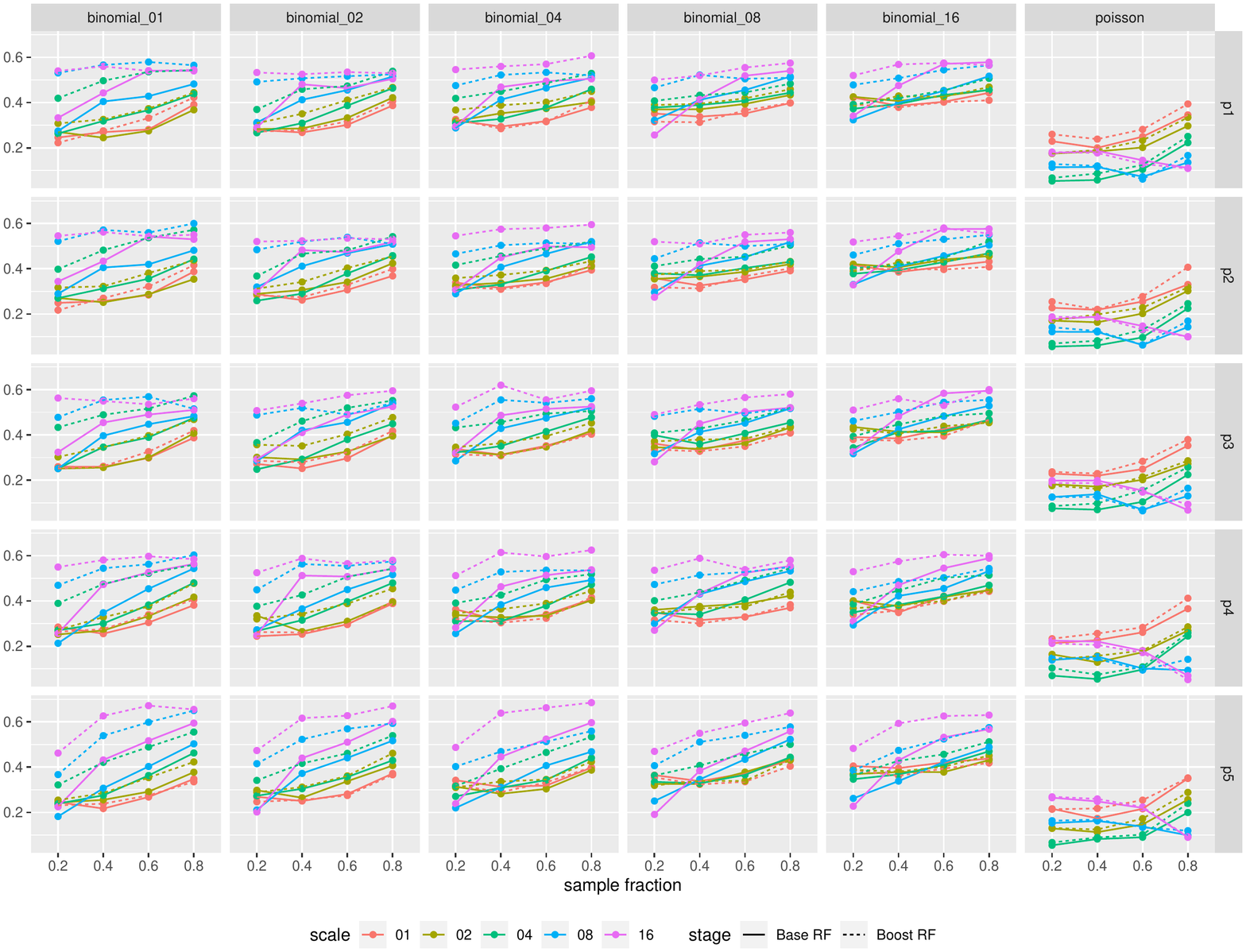}
		\end{sidewaysfigure}
		
		\item The mean squared error is predictably  smaller after the boosted stage compared to the base stage, but it also decreases as the points move away from the origin. It also increases with scale except for the binomial cases in the response space. It also decreases slightly with the subsample fraction but the number of trials in the binomial case do not seem to have any effect.
		
		\begin{sidewaysfigure}[ht]
			\centering
			\caption{Average MSE in the link space (log scale)}
			\includegraphics[width=\textheight]{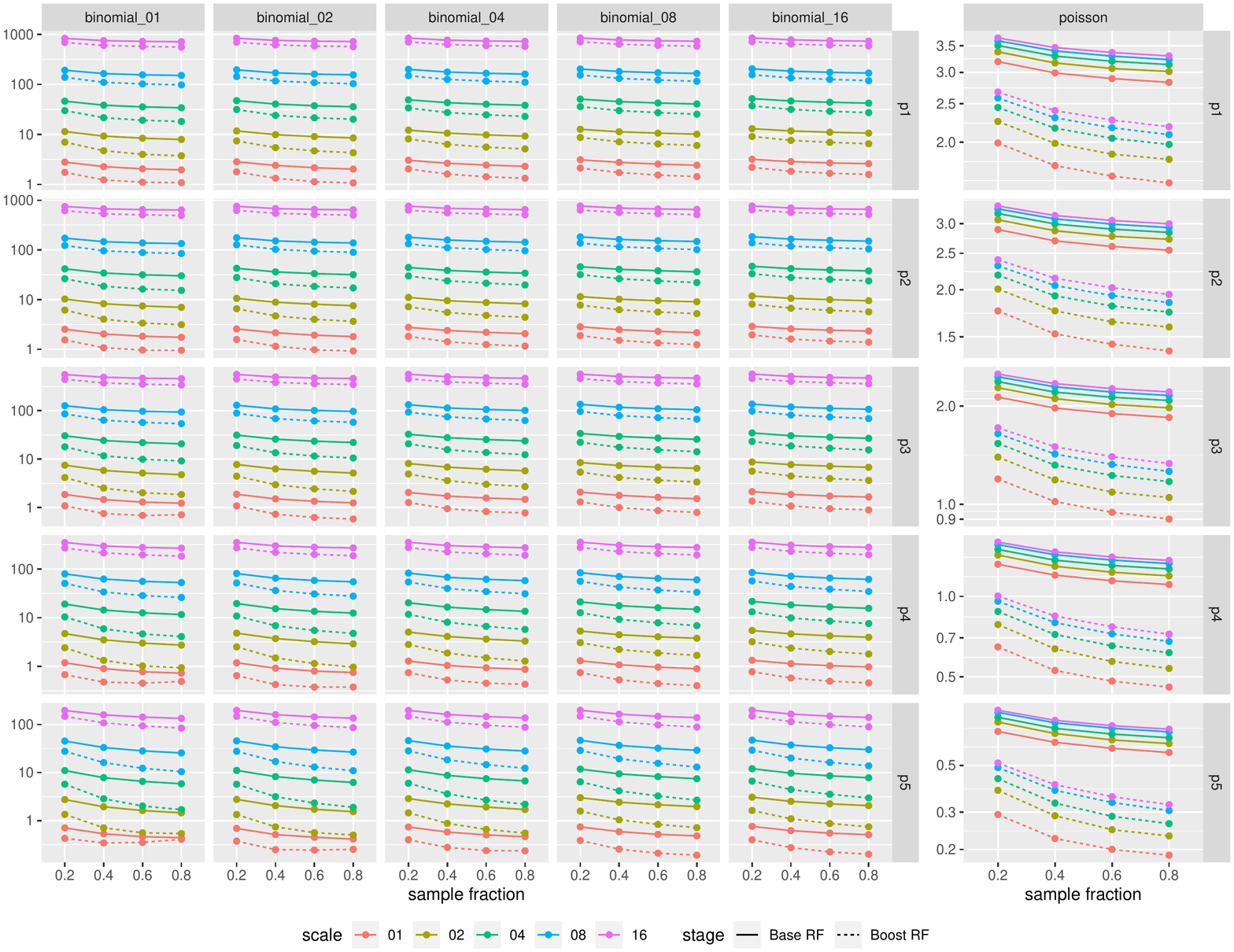}
		\end{sidewaysfigure}
		
		\begin{sidewaysfigure}[ht]
			\centering
			\caption{Average MSE in the response space (log scale)}
			\includegraphics[width=\textheight]{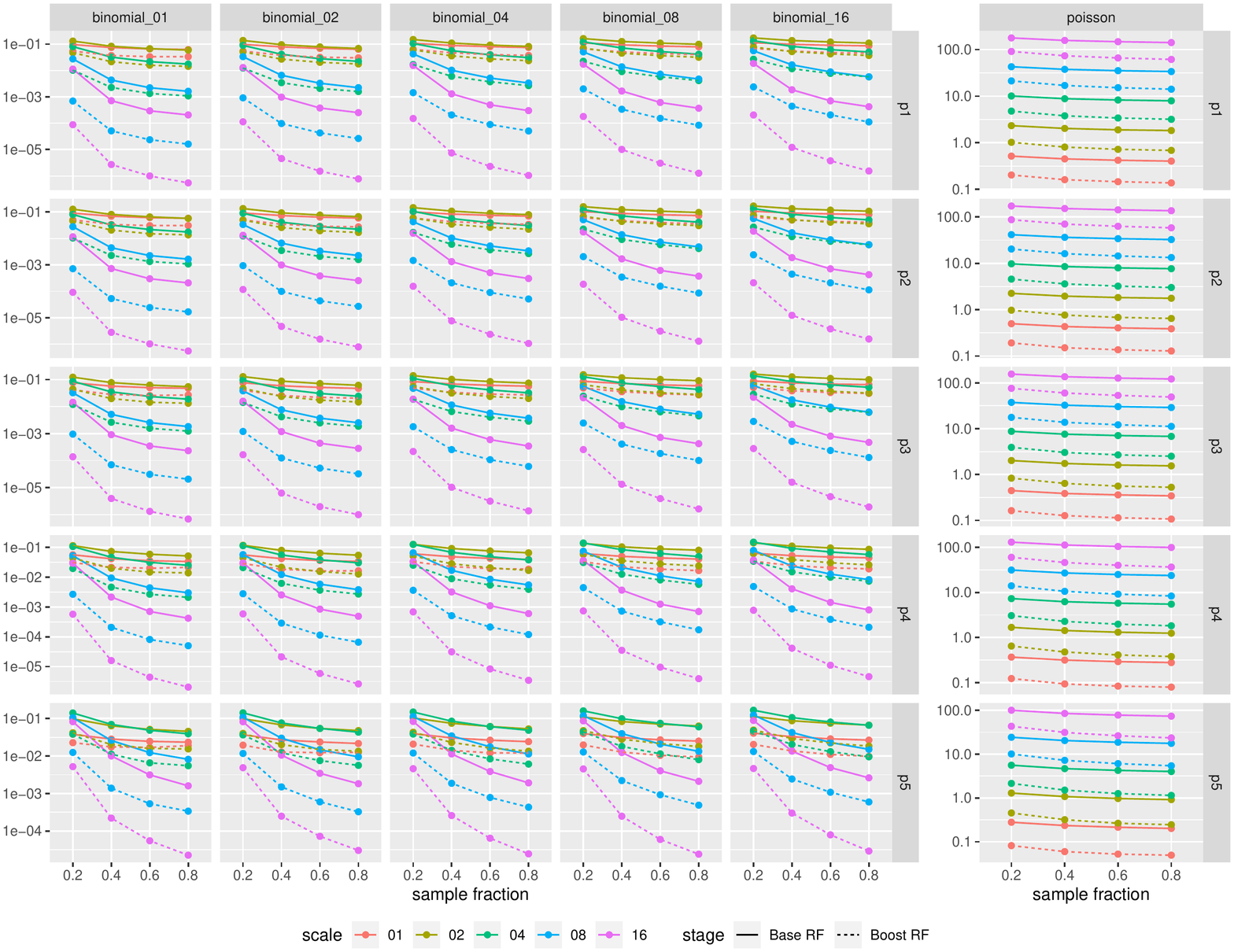}
		\end{sidewaysfigure}
	\end{itemize}

\end{document}